\documentclass[twocolumn, apj]{openjournal}

\usepackage{natbib}
\usepackage{amsmath,amssymb,verbatim}
\usepackage{graphicx}
\usepackage{hyperref}
\usepackage{txfonts}
\usepackage{xcolor}

\begin{document}


\title{How similar are narrow-line Seyfert 1 galaxies and high-z type 1 AGN?}
\author{M. Berton$^1$, E. J\"arvel\"a$^2$, A. Tortosa$^3$, C. Mazzucchelli$^4$ }

\affil{$^1$European Southern Observatory (ESO), Alonso de C\`ordova 3107, Casilla 19, Santiago, Chile
\email{marco.berton@eso.org}}
\affil{$^2$Department of Physics and Astronomy, Texas Tech University, Box 1051, Lubbock, TX 79409-1051, USA}
\affil{$^3$INAF - Osservatorio Astronomico di Roma, Via Frascati 33, I-00040 Monte Porzio Catone, Italy}
\affil{$^4$Instituto de Estudios Astrof\`isicos, Facultad de Ingenier\`ia y Ciencias, Universidad Diego Portales, Avenida Ejercito Libertador 441, Santiago, Chile}

\begin{abstract}
The recent observations of highly accreting supermassive black holes (SMBH) at very high redshift ($z>$4) with the James Webb Space Telescope (\textit{JWST}) allowed us to shed light for the very first time on the early evolutionary phases of active galactic nuclei (AGN). Perhaps unsurprisingly, several of the physical properties observed in these new objects, including those known as little red dots (LRDs), are closely reminiscent of the low-mass and high-Eddington AGN in the local Universe, and in particular of the class of narrow-line Seyfert 1 (NLS1) galaxies. However, some differences also emerged, likely due to the radically different evolutionary path and the environment where LRDs and NLS1s live. In this work, we review the multiwavelength properties of local NLS1s and briefly compare them with type 1 AGN found at high-$z$, showing that despite some differences, the study of NLS1s can be extremely useful to better understand the extreme accretion physics of high-$z$ quasars and the early stages of AGN evolution. The goal of this review is to provide the high-z community a handbook to compare their new sources with those already known in the local Universe, and to give the low-z community a starting point to venture in the early phases of AGN evolution. 
\end{abstract}

\maketitle

\newcommand{\kms}{km s$^{-1}$}
\newcommand{\ergs}{erg s$^{-1}$}
\newcommand{\chired}{$\chi^2_\nu$}
\newcommand{\hb}{H$\beta$}
\newcommand{\logxi}{$\log(\xi / \rm erg\, \rm s^{-1}\,\rm cm)$ }

\newcommand*\red[1]{\textcolor{red}{#1}}

\section{Introduction}
The launch of the James Webb Space Telescope (\textit{JWST}) has opened new horizons in the field of active galactic nuclei (AGN). It has long been known that high-$z$ quasars must accrete close to or above the Eddington ratio, that is defined as 
\begin{equation}
    \epsilon = \frac{L_{\rm bol}}{L_{\rm Edd}} = \frac{L_{\rm bol}}{1.3\times10^{38} (M/M_\odot)} \; , 
    \label{eq:edd}
\end{equation}
where $L_{\rm bol}$ is the bolometric luminosity of the AGN, and $L_{\rm Edd}$ is the Eddington luminosity of the accreting black hole of mass $M$ (in solar masses). Throughout this work, ``high-Eddington'' refers to a large luminosity-based Eddington ratio, whereas ``super-Eddington accretion'' refers to a mass accretion rate above the Eddington limit. The two terms are therefore not interchangeable, since photon trapping can keep $L_{\rm bol}/L_{\rm Edd}$ close to unity even for a super-Eddington mass accretion rate (see Sect.~3.3). Pre-\textit{JWST}, more than 500 luminous quasars were discovered at the edge of the epoch of reionization, at z$>$5.2 \citep[see][for a recent review]{Fan23} and up to z$\sim$7.5 \citep{Banados18, Yang20, Wang21}. Most of these sources harbor extremely massive black holes ($>10^9$ M$_\odot$, \citealp{Mazzucchelli23}), whose masses are hard to build in a short time. Several studies invoke either very massive seeds, which are extremely difficult to form, or a super-Eddington accretion that allows the black hole to rapidly grow to the observed masses \citep[e.g.,][]{Inayoshi20}. In this latter case, an obstacle for black hole growth is the conservation of angular momentum: accreting matter needs to lose several orders of magnitude in angular momentum to efficiently feed the black hole. However, the production of relativistic jets from these high-Eddington objects can help solve this conundrum. Jets, indeed, can drain the angular momentum, allowing a significantly more efficient accretion onto the central object \citep{Kuncic04, Jolley08a, Jolley08b, Jolley09}. 

The \textit{JWST} has started to unveil the presence of a population of significantly less massive AGN (10$^6$-10$^7$ M$_\odot$) at very high redshift that are inferred to be accreting at moderate-to-high Eddington ratios, with only some objects close to or above the Eddington limit \citep{Harikane23, Kokorev23, Kocevski23, Matthee24, Juodzbalis26}. This population may constitute the progenitors of more massive sources, that is, an intermediate phase between the early small-mass seeds and the massive quasars that were already observed in the past \citep{Inayoshi25}. Recently, some authors noticed that most of these type 1 AGN identified by the \textit{JWST}, characterized by a low black hole mass and a high Eddington ratio, formally obey the definition of narrow-line Seyfert 1 (NLS1) galaxies, a well-known class of low-$z$ AGN \citep{Maiolino25} with samples as large as 20,000 objects \citep[][see Sect.~3]{Paliya24}. 

Since their first classification based on the optical spectrum (they are type 1 AGN with FWHM(H$\beta$ $<$ 2000 \kms, \citealp{Osterbrock85}), NLS1s have shown some unique properties that are reminiscent of those of the newly discovered high-$z$ type 1 AGN. They are characterized by a low-mass black hole (10$^6$-10$^8$ M$_\odot$, \citealp{Peterson11, Cracco16, Rakshit17a, Chen18, Paliya24}) that often accrete close or above the Eddington ratio \citep{Tortosa22, Tortosa23, Paliya24}, with a steep and highly variable X-ray spectrum \citep{Boller96, Leighly99a, Leighly99b, Gallo18}, and a late-type host galaxy dominated by secular evolution \citep{Orbandexivry11, Olguiniglesias20}. Some NLS1s ($\sim$7\%, \citealp{Komossa06}) also harbor powerful relativistic jets capable of producing gamma-rays \citep{Abdo09a, Abdo09c, Foschini15, Berton15a, Paliya19a}. Because of these and many other properties, non-jetted and jetted NLS1s are both considered AGN in an early evolutionary stage \citep{Mathur00, Berton17}. If this interpretation is correct, they could represent an essential piece of the puzzle to fully explain what we are now observing in the early Universe. 

In this review, we will summarize the most important properties of NLS1s at different wavelengths. The aim of this work is to provide an overview of these sources for the non-initiated, to facilitate any future comparison with the various populations of sources recently discovered by \textit{JWST}. At the same time, in the last section, we will discuss a few recent papers that found some similarities between NLS1s and high-$z$ objects, which could prove useful for the community working on low-$z$ AGN. This work is structured as follows: in Section 2 we describe the radio properties of NLS1s with a special focus on relativistic jets; in Section 3 we examine their ultraviolet, optical, and near-infrared (NIR) properties, focusing also on the central engine properties and their host galaxy; in Section 4 we discuss their X-ray emission; in Section 5 we describe their gamma-ray spectrum; in Section 6 we summarize our current view of NLS1s, and in Section 7 we compare them with what is currently known about the new populations of broad-line AGN at high redshift. In Section 8 we provide a brief summary of our results. 
 
\section{Radio}
\label{sec:radio}
Over the whole electromagnetic spectrum, the properties of NLS1s are the most diverse in radio bands. As Seyfert galaxies with lightweight supermassive black holes, NLS1s were not expected to be a significant source of radio emission \citep{Ulvestad95}. This idea changed as larger samples of NLS1s were studied and a subset of them with strong radio emission (see Sect.~2.2 for the conventional radio-loudness definition) started to emerge \citep{Siebert99, Zhou02, Komossa06,Yuan08}. Moreover, the radio properties of these NLS1s resembled those of blazars, exhibiting dominant, variable radio emission, flat radio spectra, and compact VLBI cores with high brightness temperatures. Thus, it was not a total surprise when the first NLS1, PMN J0948+0022, was detected in gamma-rays \citep[see Sect.~\ref{sec:gamma},][]{Abdo09a}, irrefutably proving the presence of relativistic jets in this source. The discovery of relativistic jets in NLS1s expanded their radio diversity from so far undetected sub-mJy sources to those capable of hosting powerful relativistic plasma jets and exhibiting variable Jy-level radio emission.

\subsection{Radio detectability} 

The fraction of NLS1s detectable in radio greatly depends on the sample, the frequency, and the survey. Among the first NLS1 samples, the 1.4~GHz detectability using FIRST and/or NVSS varied from $\sim$7\% to $\sim30\%$, with the fraction of putatively jetted sources being a few per cent \citep{Komossa06, Zhou06}. Recently, two large NLS1 samples were published, reporting radio detection fractions of $\sim$5\% \citep[$\sim$11000 sources,][]{Rakshit17a} and $\sim$3\% \citep[$\sim$22000 sources][]{Paliya24} at 1.4~GHz using FIRST data (mean rms 0.15~mJy beam$^{-1}$, resolution 5"). \citet{Paliya24} suggest that the difference is caused by a larger fraction of higher-$z$ ($0.5 \lesssim z \lesssim 1$) NLS1s in their sample. Whereas redshift likely plays a role, so does the quality of the sample. When the optical spectra of the $\sim$11000 sources identified in \citet{Rakshit17a} were fitted manually and the results checked one by one, only $\sim$4000 sources could be reliably classified as NLS1s \citet{Berton20a}. The average redshift of this sample, deemed the 4k sample, is significantly lower than that of the original sample, indicating that the higher the redshift the more unreliable the classification. Thus, the \citet{Rakshit17a} and \citet{Paliya24} samples might include a large fraction of contaminating sources that can have a major impact on the population statistics. The FIRST radio detection rate among the 4k sample is 8\% \citep{Varglund25}. They also studied the detection rate at 144~MHz (LOFAR) and 3~GHz (VLASS Epochs 1 and 2). Probably as a result of better sensitivity (median rms 83~$\mu$Jy beam$^{-1}$), lower frequency (most NLS1s have steep spectra and star formation activities are more prominent), and slightly lower resolution (6", less emission resolved out), the detection rate of NLS1s at 144~MHz is 44.6\%. At 3~GHz (mean rms of 120 $\mu$Jy beam${-1}$, resolution 2.5"), the detection rate decreases to $\sim$5.3\%, likely due to steep radio spectra and additional resolved-out emission. Future radio facilities, such as the next generation VLA (ngVLA) and the Square Kilometre Array Observatory (SKAO), with unprecedented sensitivities are likely to detect an increasing number of NLS1s, though at low flux densities it will also become increasingly hard to determine whether the radio emission originates from the AGN or (circumnuclear) star formation, a common phenomenon in NLS1s. 

\subsection{Origin of radio emission}

For decades, the radio loudness parameter \citep[$RL$, defined as $f_{\nu}(5\,{\rm GHz})/f_{\nu}(B)$, ][]{Kellermann89} was used to divide AGN into radio-loud ($RL > 10$) and radio-quiet ($RL < 10$) sources. The distribution was believed to be bimodal \citep{Cirasuolo03} and reflect the origin of the radio emission in these sources: the radio-loud sources were associated with AGN hosting powerful radio jets, and the radio-quiet with non-jetted AGN. However, the dichotomy turned out to be an observational bias \citep[also within the NLS1 class, e.g.,][]{Jarvela15} and with the emergence of the radio loudness continuum, the parameter lost most of its predictive power. Indeed, it was later suggested that AGN should be classified based on their actual physical properties, rather than a somewhat arbitrary parameter \citep{Padovani17, Jarvela17, Foschini17, Berton21c, Wang25}. However, in the case of NLS1s, this turned out to be challenging, especially at low frequencies ($<$ 10~GHz), as several phenomena producing radio emission, such as radio jets of varying power and brightness, outflows, and star formation, can co-exist, and distinguishing the contribution of each can be tricky. Broadly speaking, NLS1s can be divided into host-dominated and AGN-dominated sources, with most individuals somewhere between the two extremes \citep{Jarvela22}. 

\begin{figure}
    \centering
    \includegraphics[width=0.48\textwidth]{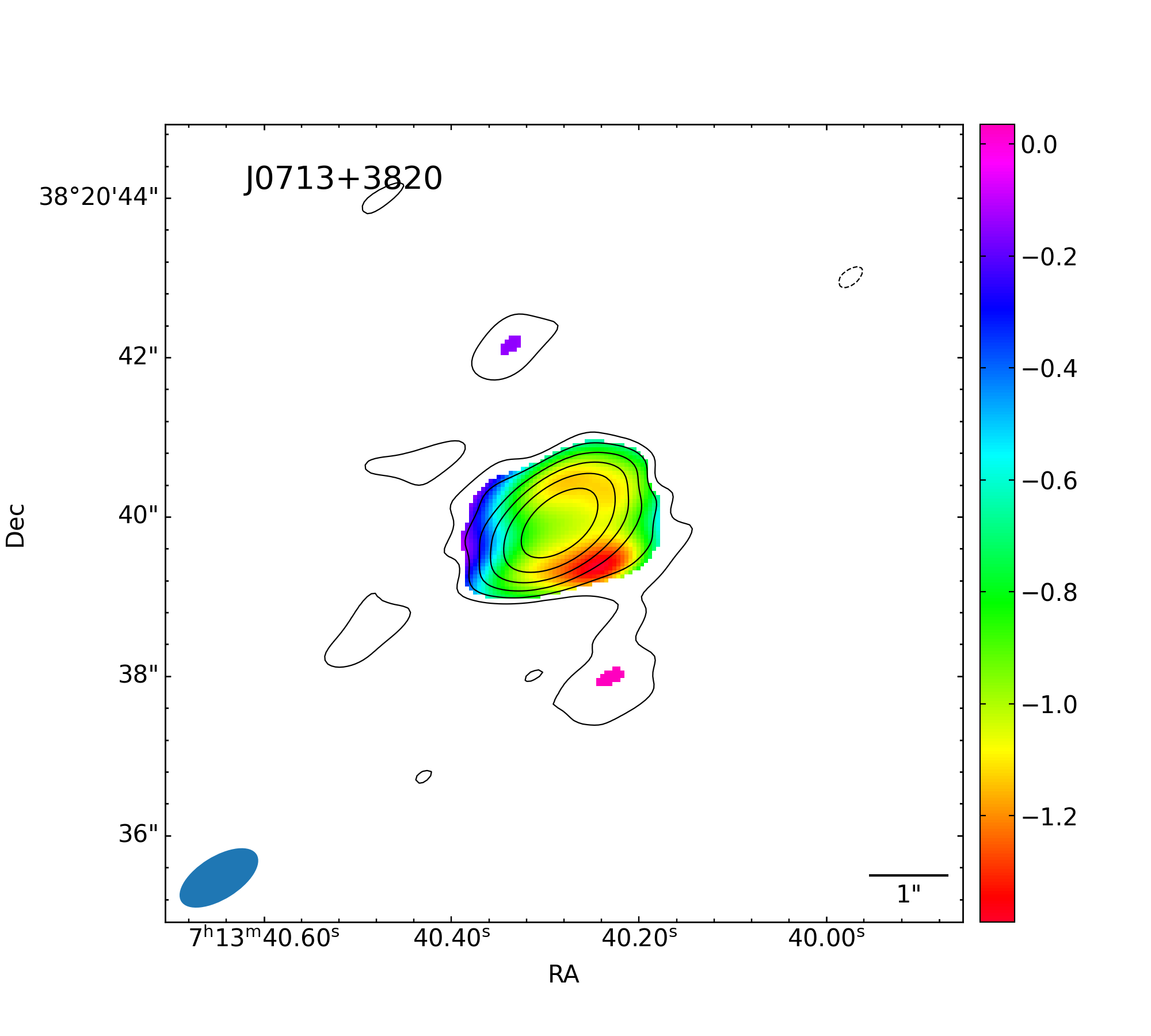}
    \caption{5.2~GHz radio map and spectral index map of J0713+3820, showing patchy radio morphology indicating the presence of circumnuclear star formation. The rms = 13~$\mu$Jy beam$^{-1}$, the contour levels are at -3, 3, 6, 12, 24, and 48 $\sigma$, and the beam size is 2.45~kpc $\times$ 1.19~kpc. Image credit: \citet{Jarvela22}.}
    \label{fig:J0713-sf}
\end{figure}

In host-dominated NLS1s, the often patchy radio morphology, as seen in Fig.~\ref{fig:J0713-sf}, sometimes tracing the host galaxy features, indicates star formation as the main source of radio emission. These NLS1s generally have low luminosities and, based on MIR/FIR - radio -relations for star formation activity, no significant AGN contribution is needed to explain the measured radio flux density \citep{Jarvela22, Chen22}. However, highlighting the peculiar nature of NLS1s, \citet{Caccianiga15} identified a subset of NLS1s that can be classified as radio-loud but in which the radio emission seems to be produced by starburst-level star-forming activity. Generally, star formation regions exhibit steep spectral indices ($\alpha \sim-0.7$, $S_{\nu} \propto \nu^{\alpha}$) and their contribution quickly diminishes towards higher frequencies.

A large fraction of NLS1s show radio properties that are characteristic of both circumnuclear star formation and AGN activity. They can show, for example, a flat-spectrum core, variability, a high radio luminosity, and/or conical structures resembling a jet or an outflow, and on the other hand, patchy radio emission, and a strong indication of star formation in other bands, such as polycyclic aromatic hydrocarbon (PAH) emission lines. In these sources, the radio emission is produced by the AGN and the star formation activities, with varying levels of contribution. These can be distinguished by using spectral, morphological, luminosity, and brightness temperature information, often requiring mas-scale observations to reliably determine the contribution of each. 
 
In the other extreme of the continuum reside NLS1s whose radio emission is dominated by the nucleus. These sources show significant diversity, spanning from sources with poorly collimated jets or wide-angle outflows to sources showing powerful collimated pc- to kpc-scale jets. Poorly collimated jets and outflows are low-luminosity, but can be distinguished from star formation based on their asymmetric morphologies. However, jets and outflows can also co-exist, further complicating the picture \citep{Gu10, Longinotti18}. Though usually in these cases, the jet dominates the overall radio emission.

Collimated jets in NLS1s display varying levels of power, possibly determined by a combination of the black hole mass and spin, and the accretion disk state \citep{Chamani21}. A small fraction of NLS1s are capable of hosting powerful relativistic jets \citep{Richards15, Lister16}, comparable to those of flat-spectrum radio quasars (FSRQ) when scaled with the black hole mass \citep{Heinz03, Foschini15}. Based on VLBI observations, some of these jets show highly superluminal motion (up to $\sim$10$c$) and Lorentz factors around $\sim$10, implying small viewing angles. Also, the polarimetric properties of NLS1s are similar to those of FSRQs, further strengthening the connection between the two classes \citep{Takamura23}. As expected, in single-dish multifrequency observations, jetted NLS1s exhibit intense variability and blazar-like flaring behavior \citep{Angelakis15, Lahteenmaki17}. 

\begin{figure}
    \centering
    \includegraphics[width=0.48\textwidth]{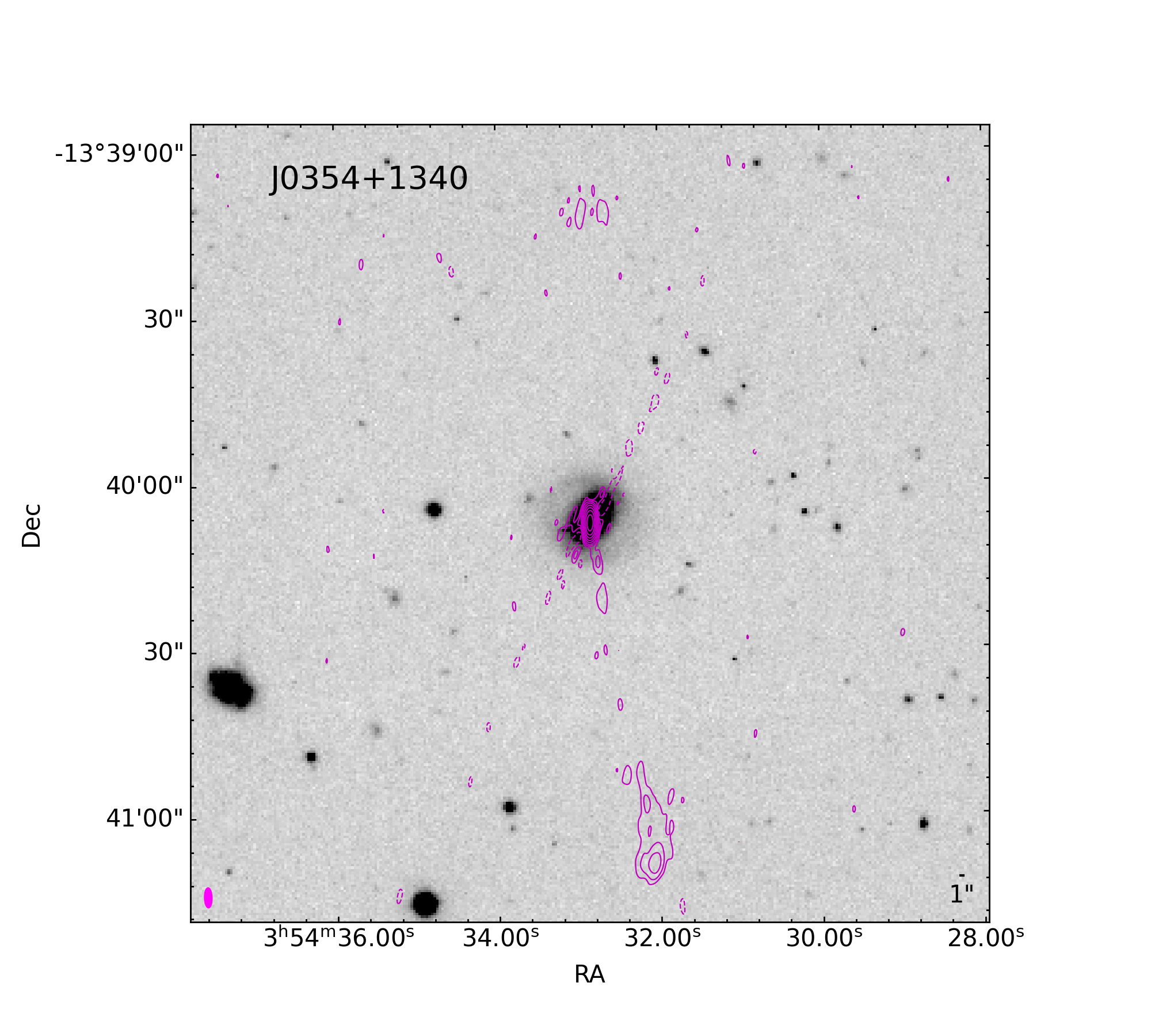}
    \caption{5.5~GHz radio map of J0354$-$1340 overlaid with the $J$-band host galaxy image. The southern jet has a projected size of 93.5~kpc, and an estimated deprojected size of 127~kpc. The rms = 8~$\mu$Jy beam$^{-1}$, the contour levels are at $-$3, 3, 6, 12, 24, 48, 96, and 192 $\sigma$, and the beam size is 5.25~kpc $\times$ 1.93~kpc. Image credit: \citet{Vietri22}.}
    \label{fig:J0354-largejet}
\end{figure}

\begin{figure}
    \centering
    \includegraphics[width=0.48\textwidth]{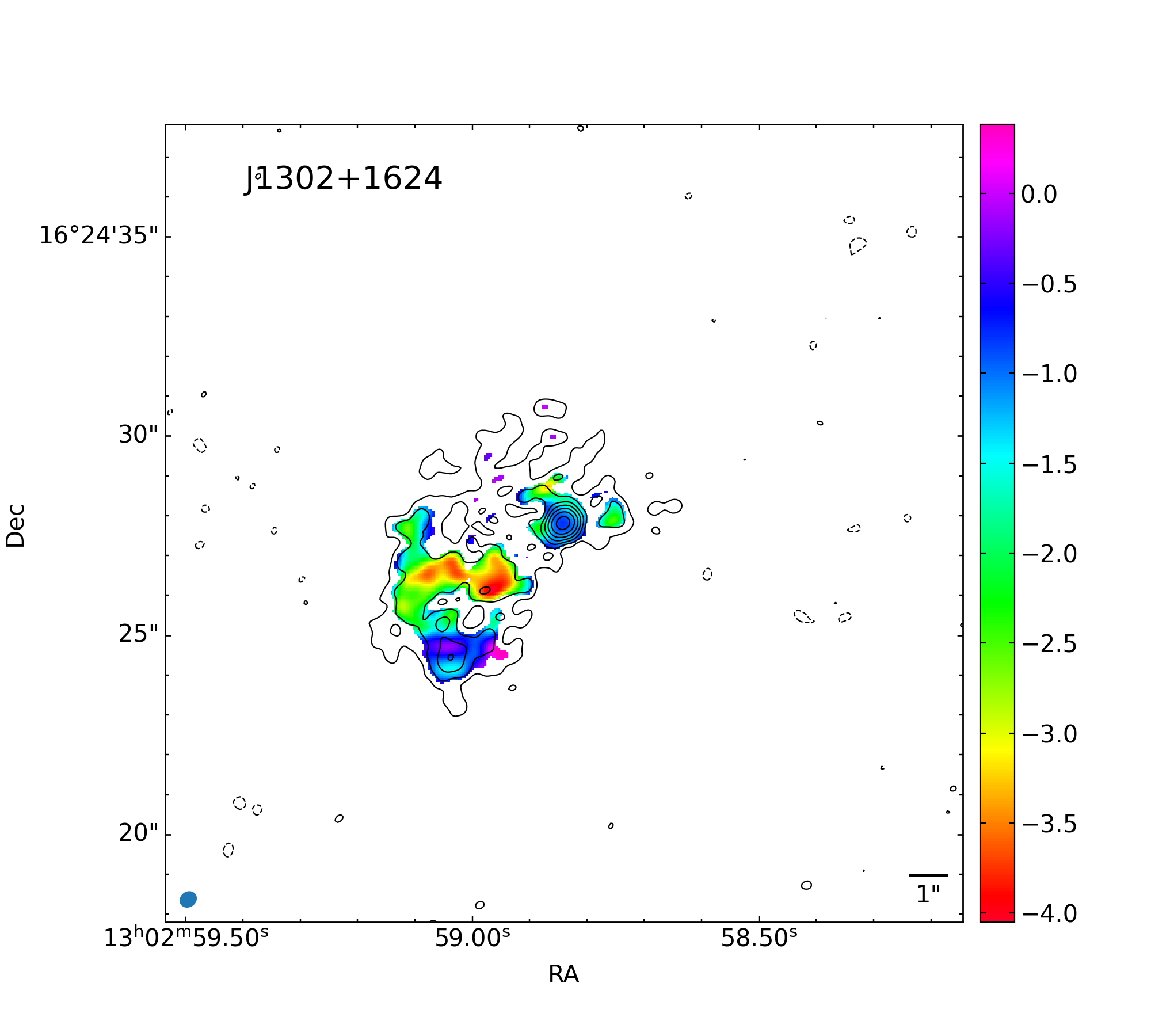}
    \caption{5.2~GHz radio map and spectral index map of J1302+1624 (Mrk 783). A region with a very steep spectral index is seen south-east of the nucleus, whereas the currently active jet is directed towards north-east \citep{Congiu20}. The rms = 12~$\mu$Jy beam$^{-1}$, the contour levels are at $-$3, 3, 6, 12, 24, 48, and 96 $\sigma$, and the beam size is 0.58~kpc $\times$ 0.50~kpc. Image credit: \citet{Jarvela22}.}
    \label{fig:Mrk783}
\end{figure}

Also, some NLS1s with large jets seen at larger viewing angles have been discovered \citep[Fig.~\ref{fig:J0354-largejet} and e.g.,][]{Vietri22, Jarvela22, Chen24}. These jets are estimated to have deprojected sizes around a hundred kpc, indicating that the jets in NLS1s can be quite a long-lived phenomenon. So far, clear signs of relic radio emission have been found only in one NLS1 \citep[Fig.~\ref{fig:Mrk783} and ][]{Congiu17}, and even in this case, it is unclear whether the jet actually turned off or just changed direction.

\begin{figure}
    \centering
    \includegraphics[width=0.48\textwidth]{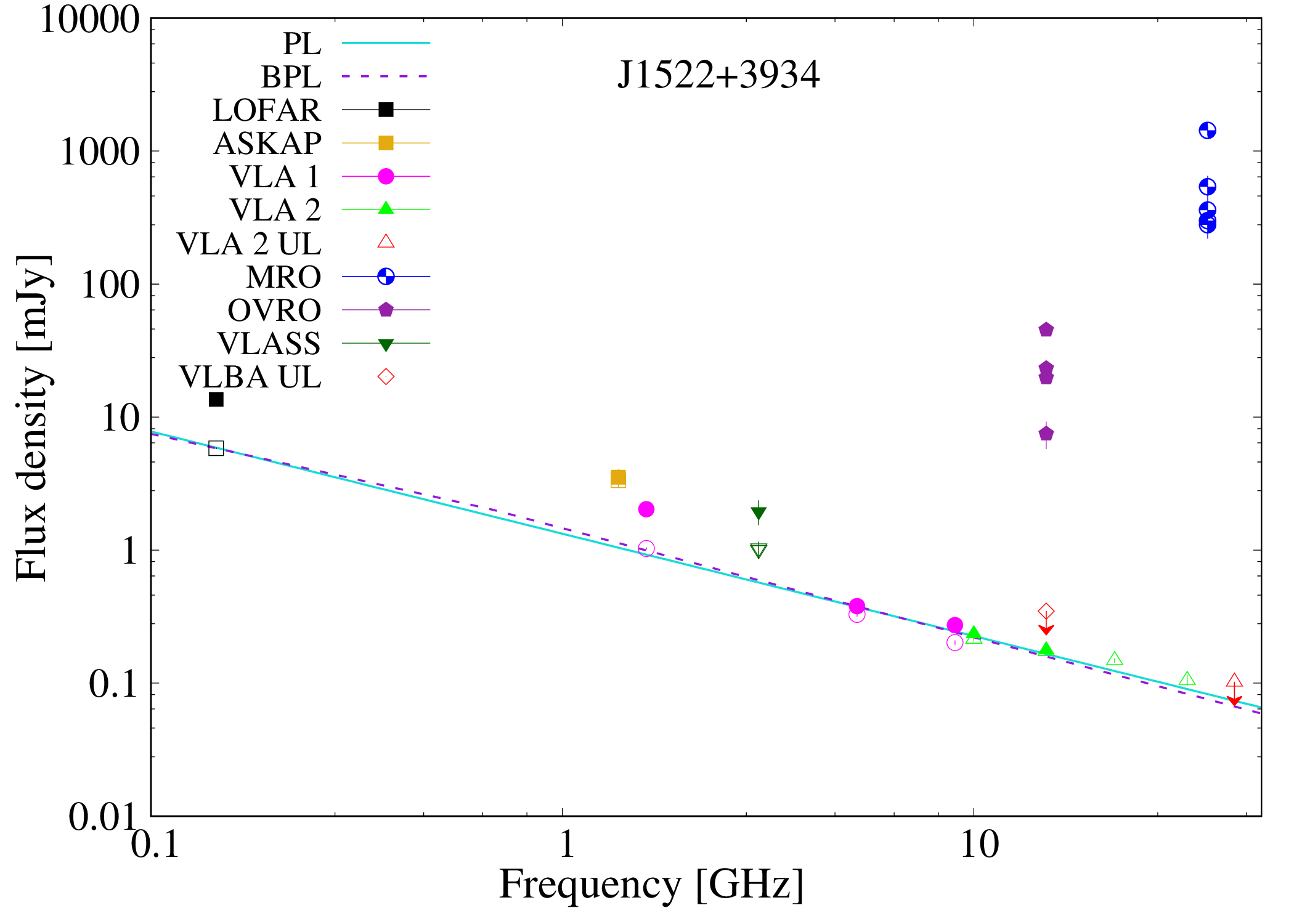}
    \caption{Non-simultaneous radio spectrum of J1522+3934. Symbols and colors are explained in the legend. Filled symbols denote integrated flux densities and empty symbols mark peak flux densities, except empty red symbols with downward arrows that are used for upper limits. VLA 1 data from \citet{Berton20b} and VLA 2 data from \citet{Jarvela24}. The power law (PL) and broken power law (BPL) fits shown are the fits to the peak flux densities. Image credit: \citet{Jarvela24}.}
    \label{fig:J1522-spectrum}
\end{figure}

\subsection{Extreme radio variability} 

Metsähovi Radio Observatory (Finland) specializes in long-term monitoring of large samples of AGN at 37~GHz, including also $\sim$250 NLS1s \citep{Lahteenmaki17}. This monitoring program led to the discovery of seven NLS1s previously classified as radio-quiet or with no known radio emission that undergo recurring Jy-level flares at 37~GHz \citep{Lahteenmaki18}. Three of these sources have also been detected at 15~GHz by the Owens Valley Radio Observatory (OVRO), showing flux densities of a few tens of mJy. The radio flares in these NLS1s resemble those seen in blazars but exhibit extremely high-amplitude variability (three to four orders of magnitude) at very short timescales ($e$-folding timescales of days to a few weeks) \citep{Jarvela24}. In the low activity state, as seen in Fig.~\ref{fig:J1522-spectrum}, the radio spectrum shows a spectral index around $-$0.7, characteristic to optically thin radio emission, either from the AGN or circumnuclear star formation. In the high state, however, the spectrum seems to be very inverted, though it should be noted that the 15 and 37~GHz data are not simultaneous. Intriguingly, no radio jets were detected in these sources at 15~GHz using the Very Long Baseline Array. 

The apparent absence of jets, coupled with the amplitude and temporal variability, sets strict constraints on the hypotheses attempting to explain the behavior seen in these NLS1s. \citet{Jarvela24} were able to rule out the most common intrinsic and external phenomena causing variability in AGN, including ordinary intrinsic jet variability, stationary synchrotron self-absorption or free-free absorption, extreme scattering events, and microlensing, and narrowed the alternatives down to a few options. 1) First, interaction between a jet or a jet base and broad-line region (BLR) clouds or a star can lead to shocks that efficiently accelerate electrons, leading to enhanced flux densities \citep{Fraixburnet92, Gomez00, Delpalacio19}. Based on simulations, these interactions result in flares lasting from less than a day to a few days, and thus match the timescales of these flares. However, these interactions should be multiwavelength events, which have not been observed in the NLS1 flares. 2) The second hypothesis could explain the flares as an interplay between a small-scale relativistic jet and ionized BLR clouds. In this scenario, the jet is still confined within the ionized BLR, and its emission is totally free-free absorbed most of the time. The flares happen when there is a gap in the BLR cloud coverage and the nucleus is temporarily revealed. Indeed, some indication of this kind of behavior was found in one of these sources. Its X-ray spectrum in the low state shows a possible warm absorber, which disappears in a spectrum taken right after a 37~GHz flare \citep[][L\"ahteenm\"aki et al., in prep]{Romano23}. 3) The third alternative is magnetic reconnection in the jet or in the black hole magnetosphere \citep{Degouveiadalpino10, Kadowaki15, Ghisellini08, Nalewajko11}. It has been observed to take place in the jet, and the timescales of these events at high energies can range from minutes to days, but not much is known about the observable radio properties. Due to the lack of jets in these NLS1s, magnetic reconnection in the black hole magnetosphere is a more appealing option, but direct evidence of that is still totally missing. However, it has been suggested that this mechanism is responsible for the radio and gamma-ray emission in low-luminosity AGN \citep{Degouveiadalpino10, Kadowaki15}. 

Additional studies carried out on these sources have not been conclusive \citep{Ojha24, Crepaldi25}. The possibility that the extreme variability is due to something so far unknown cannot be ruled out either, and naturally, the variability mechanism does not need to be the same in all sources. Either way, it looks like these sources are exhibiting a new kind of AGN variability, repercussions of which remain uncharted.
\begin{figure}[!t]
    \centering
    \includegraphics[width=0.48\textwidth]{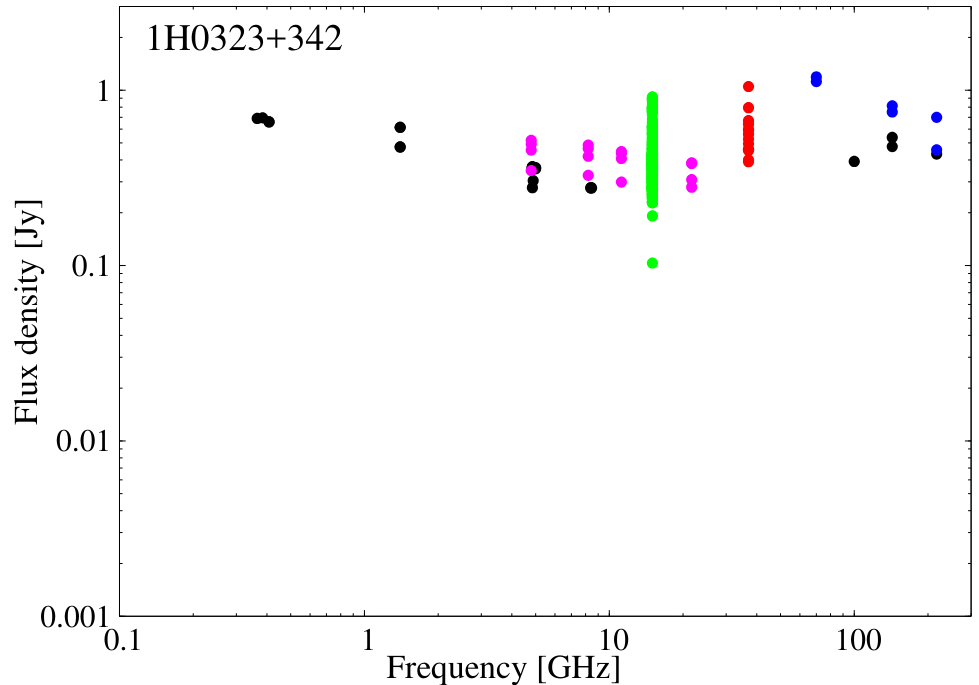}
    \caption{Non-simultaneous radio spectrum of 1H 0323+342. Black denotes archival data, magenta RATAN-600 data, green OVRO data, blue \textit{Planck} data, and red MRO data. Image credit: \citet{Lahteenmaki17}.}
    \label{fig:1H0323-spectrum}
\end{figure}

\begin{figure*}[!ht]
    \centering
    \includegraphics[width=0.95\textwidth]{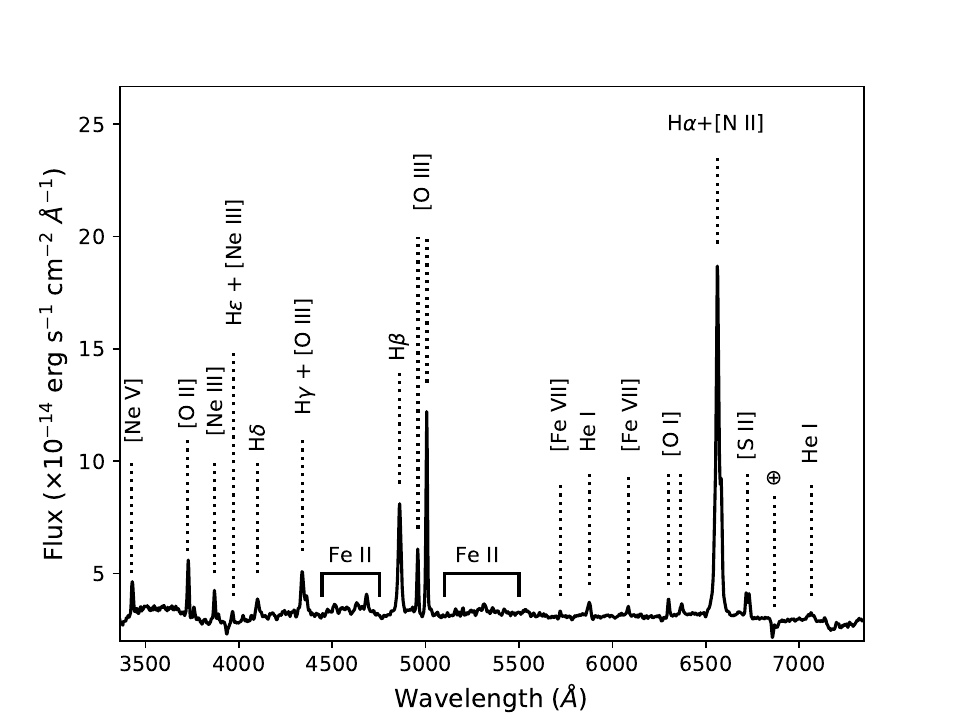}
    \caption{Optical spectrum of NGC 4051, the closest NLS1 (16.6 $\pm$ 0.3 Mpc, \citealt{Yuan21}). The main emission lines are highlighted. The symbol $\oplus$ denotes a telluric absorption. The spectrum was obtained between 2014-03-30 and 2014-04-01 with the Boller \& Chivens spectrograph of the Asiago 1.22m telescope. The exposure time is 12000s, the spectral resolution is R$\sim$700.}
    \label{fig:opt_spectrum}
\end{figure*}
\subsection{(Sub-)mm view and radio line emission}

Radio continuum observations of NLS1s above $\sim$50~GHz and into the sub-mm regime are rare. The only survey at these frequencies was performed by \textit{Planck} satellite, which surveyed the entire sky in nine bands spanning from 30 to 857~GHz \citep{Planck16}. Only a few NLS1s, all hosting powerful relativistic jets, were detected by \textit{Planck} \citep{Lahteenmaki17}. These sources show variable, but in general steep-spectrum radio emission at \textit{Planck} frequencies, as seen in Fig.~\ref{fig:1H0323-spectrum}. The only targeted study is a monitoring campaign of a small sample of jetted NLS1s at 2.64-142.33~GHz using the Effelsberg 100-m Radio Telescope and the Institut de radioastronomie millimétrique (IRAM) 30m radio telescope \citep{Angelakis15}. As expected, these sources were found to exhibit highly variable, blazar-like emission and are clearly dominated by their relativistic jets.

Measurements of the total CO line emission provide molecular gas masses, compared with the star formation rates, show that NLS1s generally lie above the galaxy main sequence and have short molecular-gas depletion times, as further discussed in Sect.\ref{sec:host} \citep{Salome23}. To the best of our knowledge, only one NLS1, IRAS 17020+4544, has spatially resolved observations of any radio emission lines \citep{Longinotti18, Salome21}. This source was observed with the NOrthern Extended Millimeter Array (NOEMA) utilizing the CO(1-0) line and shows biconical kpc-scale molecular outflows \citep{Longinotti23}. Combining these results with an earlier detected ultrafast outflow (UFO) (see Sect.~4) allowed the authors to conclude that IRAS 17020+4544 exhibits an energy-conserving galaxy-scale outflow and is thus a prime example of AGN feedback in action.

\section{UV/optical/NIR}
The original definition of NLS1s is based on optical spectroscopy (see Sect.~\ref{sec:properties}), and the most prominent emission lines from ionized gas can be identified. In the NIR, instead, the nuclear continuum can include emission from the innermost hot dust near the sublimation radius, while at lower AGN luminosities the stellar component of the host galaxy can contribute substantially \citep{Barvainis87, Fischer06, Jarvela18}.

In the Southern sky, due to the lack of deep spectroscopic surveys, a relatively small number of NLS1s is known ($\sim$200, \citealp{Chen18, Durre22}). In the North, instead, after the advent of the Sloan Digital Sky Survey (SDSS, \citealp{York00}), several surveys have been dedicated to identifying NLS1s, most of them using single-epoch spectroscopy, finding more than 20,000 objects in the Northern hemisphere only \citep{Zhou02, Zhou06, Cracco16, Rakshit17a, Paliya24}. In several cases, however, automatic procedures were used to identify and analyze the NLS1 spectra. As already pointed out in Sect.~\ref{sec:radio}, this can lead to misleading results. Spectra, especially when the signal-to-noise ratio is low, can be very hard to study without human supervision, and often other classes of AGN, such as broad-line Seyfert 1s (BLS1s) or intermediate type Seyfert galaxies, can end up contaminating the samples \citep[e.g.,][]{Jarvela20}. Therefore, extreme care should be used when dealing with NLS1s' spectroscopic properties, especially with large samples.

\subsection{Spectral properties}
\label{sec:properties}
The original classification by \citet{Osterbrock85} was based on the optical spectra of a small sample of these objects. They noted that all of them had unusually narrow Balmer lines compared to those of other type 1 AGN, a flux ratio [O III]/H$\beta < 3$, and similar luminosity to other Seyfert 1 galaxies. Later spectropolarimetric observations by \citet{Goodrich89} proved that all of them had a full-width at half maximum of H$\beta < 2000$ \kms. Since then, this has become a hard threshold to classify NLS1s. From a physical point of view, this threshold is not very meaningful, but the definition allows us to group together all the sources with the most extreme properties. 

An example of a typical NLS1 optical spectrum is shown in Fig.~\ref{fig:opt_spectrum}. It is characterized by narrow permitted lines with a FWHM between 1000 and 2000 \kms. Their Balmer lines can typically be modeled with a Lorentzian profile \citep{Cracco16}, which is indicative of turbulent motion of the gas in the BLR \citep{Gaskell09, Kollatschny11, Goad12} or dust-driven outflow models \citep{Naddaf22}. It is worth noting that some exceptions to this rule exist. In some cases, this is only due to the poor signal-to-noise ratio of the spectra: poor quality observations can hide the wings of the Lorentzian profile in the noise, thus mimicking a Gaussian profile. In some cases, however, the Balmer lines are genuinely different, and they are reproduced with a combination of Gaussian components, typically one for the emission coming from the NLR, and two for the BLR emission \citep{Popovic04}. The two broad Gaussian components are a phenomenological description of the line profile and do not necessarily correspond to two spatially distinct BLR regions. Such Gaussian objects may have an intrinsically different geometry of the BLR, dominated by a Keplerian motion instead of turbulence \citep{Berton20a}.

Fe II multiplets are a common feature in NLS1 spectra from the UV to the NIR. These lines are produced within the BLR, typically have a FWHM comparable to that of H$\beta$. Together with the broad permitted Balmer component, their presence confirms the type 1 nature of NLS1s despite the narrow line widths; the [O III]/H$\beta < 3$ ratio is instead part of the historical classification criterion. Some NLS1 spectra, with particularly strong iron features, are used to build the Fe II templates for type 1 AGN \citep{Kovacevic10, Shapovalova12}. However, the strength of the Fe II multiplets can be very different. Usually it is measured by the parameter R4570, defined as the ratio between the integrated flux of the Fe II blend at 4434--4684~\AA\ and that of the \hb\ broad component \citep{Marziani18b}. Some sources have an R4570 $\gtrsim$ 2, while other sources have R4570 $\sim$ 0. The origin of the iron lines is still heavily debated \citep{Panda19, Marziani21, Panda21}, but it may be produced by a combination of collisional and resonance-fluorescence processes \citep{Sigut98, Verner99}. It is worth noting that Fe II strength is not a direct metallicity indicator, since it also depends on density, column density, ionization parameter, and microturbulence \citep{Panda18}.

\begin{figure}[!t]
    \centering
    \includegraphics[width=0.48\textwidth]{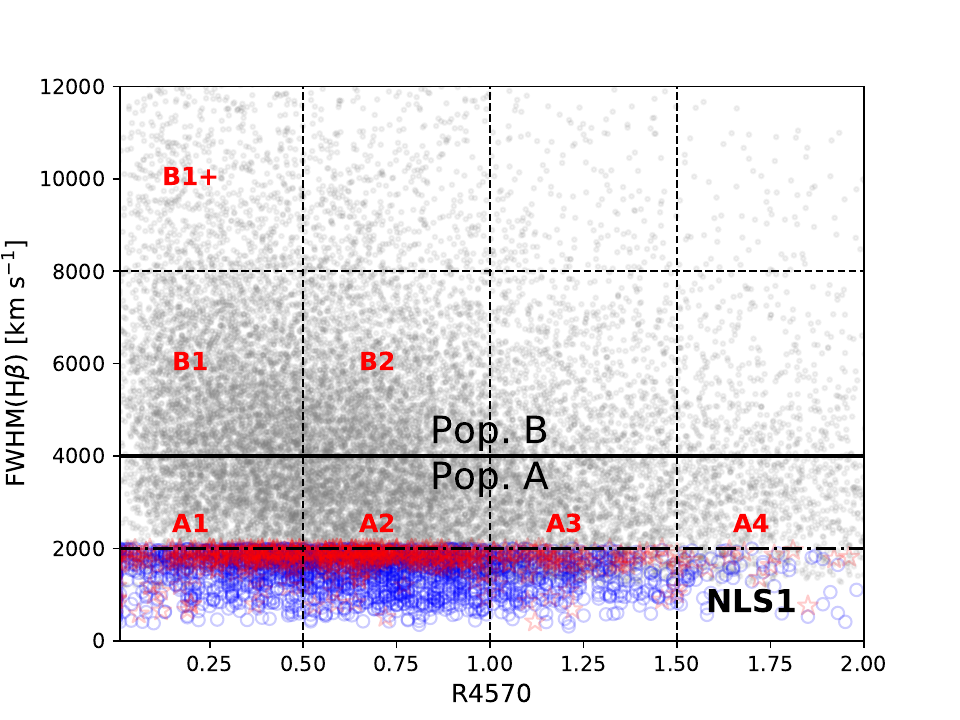}
    \caption{The R4570-FWHM(\hb) plane for type 1 AGN. The blue circles and red stars represent NLS1s analyzed in \citet{Berton20a} whose \hb\ line profile is better reproduced by a Lorentzian and a Gaussian function, respectively. The grey points come from the quasar sample of \citet{Shen11}. The red labels indicate the populations defined by \citet{Sulentic00}.}
    \label{fig:ev1}
\end{figure}
 
In general, it is worth noting that there is no sharp transition of physical properties happening around the NLS1 FWHM threshold. The AGN eigenvector 1 \citep[EV1,][]{Boroson92, Shen14}, also known as the quasar main sequence \citep{Sulentic00}, is a useful tool to graphically present the properties of type 1 AGN on the R4570-FWHM(\hb) plane (Fig.~\ref{fig:ev1}). The main sequence, indeed, shows a real discontinuity only around a FWHM of 4000 \kms, dividing AGN to two populations of objects called A and B. On one hand, population A sources show prominent Fe II multiplets, a Lorentzian profile in the permitted lines, low equivalent width (EW) of [O III]$\lambda$5007, strong outflows in the [O III] and C IV $\lambda$1549 lines, and a steep X-ray spectrum. On the other hand, population B sources have weak or no Fe II, their Balmer line can be modeled with composite Gaussian profiles, their [O III] have larger EW, they do not show any significant outflows in the forbidden or high ionization lines, and have a harder X-ray spectrum \citep{Marziani18b}. The main driver of the EV1 is thought to be the Eddington ratio which, in combination with metallicity and orientation, can explain most of the observational properties along the quasar main sequence (\citealp{Panda19, Panda23}, for a comprehensive review on the Eddington ratio in quasars, see \citealp{Marziani25}). In this framework, NLS1s represent the most extreme sources within population A, and are therefore worth investigating on their own, even if their properties are somewhat similar to those of other population A objects. The FWHM limit furthermore introduces an inevitable selection bias as soon as the \hb\ line moves in the infrared due to redshift. Large spectroscopic surveys such as SDSS are indeed limited to wavelengths lower than 9000 \AA, thus restricting our ability to find NLS1s up to z$\sim$0.85, with one of the farthest being 3C 286 exactly at this redshift \citep{Yao21, Komossa24}. Above this $z$, the FWHM of the Mg II $\lambda$2798 line has been proposed as a tool to identify NLS1s \citep{Rakshit21a}, but its relation with the FWHM \hb\ should be better calibrated. 

Outflows are also commonly observed in the emission lines of NLS1s, and most prominently in the high-ionization lines such as [O III]$\lambda\lambda$4959, 5007 and the C IV $\lambda$1549. The C IV results are obtained from dedicated UV spectroscopy rather than the optical surveys used to select the H$\beta$ samples \citep{Sulentic08}. The presence of outflows is usually correlated with the strength of the Fe II multiplets and with the presence of a blueward asymmetry in \hb, and it is also inversely correlated with the strength of the [O III] \citep{Boroson92, Sulentic00, Marziani03}. 
This is in agreement with the physical interpretation of the EV1, i.e. the Eddington ratio drives the physical properties of the source. Indeed, it was found that in the general AGN population, outflows occur more frequently in high-Eddington sources \citep{Mullaney13, Woo16}. These features indicate a bulk motion of the gas producing these emission lines, that is, the inner part of the BLR and the whole NLR. It is worth noting that a remarkable blueshift in the [O III] lines seems to be related to the presence of relativistic jets but with no preference for low-inclination sources \citep{Berton16b, Berton21a}. This is in agreement with integral-field observations of AGN with small relativistic jets, showing that a jet crossing the NLR produces outflows in every direction, and not only parallel to its propagation direction \citep{Venturi21}. A side effect of the large number of outflows is that measuring the redshift in NLS1s is not a trivial task. Many emission lines, especially in strong Fe II emitters, can be blueshifted, while absorption lines, as expected in a type 1 AGN, are basically lost in the non-thermal continuum. The best proxy for the restframe velocity is provided by low-ionization emission lines, such as the [S II] $\lambda\lambda$ 6716,6731, [O II]$\lambda$3727, or [O I]$\lambda$6300 \citep{Komossa08}. 

\subsection{Host galaxy} \label{sec:host}
Early observations of the host galaxy of nearby NLS1s revealed a very high fraction (up to 100\%) of barred spirals \citep{Krongold01, Crenshaw03, Ohta07}. The \textit{Hubble} Space Telescope confirmed this result, showing that the vast majority of NLS1s are hosted in spirals with pseudobulges \citep{Mathur12}. This supports the hypothesis that the black hole in NLS1s is fed by means of secular evolution, with the bar playing an essential role in transporting the gas from the outer regions of the galaxy to its nucleus \citep{Orbandexivry11}. This result is in agreement with studies on the large-scale environment of NLS1s, which, unlike other AGN, tend to be found either in voids or, more generally, in low-density environments \citep{Jarvela17}. Additionally, the host galaxy of NLS1s is characterized by strong circumnuclear star formation, located in the bar or rings. These star-forming regions have been identified using polycyclic aromatic hydrocarbons as tracers \citep{Sani10}, by their strong radio emission \citep{Caccianiga15} or, more recently, by subtracting the nuclear point-spread function from integral-field observations \citep{Winkel22}. 
 
The host galaxy of jetted NLS1s, instead, has been poorly studied until very recently, because most of these sources actually reside at relatively higher redshifts (z$\sim$0.5) and are therefore difficult to observe from ground-based facilities without the aid of adaptive optics. Even though some isolate studies seem to show the presence of an elliptical host for these sources, making them more similar to other jetted AGN, such as radio galaxies and blazars \citep{Dammando17, Dammando18}, the vast majority of authors agree that jetted NLS1 are hosted in late-type galaxies just like their non-jetted counterparts \citep{Olguiniglesias17, Jarvela18, Berton19a, Olguiniglesias20, Hamilton21, Vietri22, Varglund22, Varglund23}. While non-jetted NLS1s are almost ubiquitously hosted in isolated galaxies, interacting systems are more commonly observed in jetted objects \citep{Anton08, Berton19a}. Merging may be a key ingredient to launch a relativistic jet \citep{Chiaberge15}, but not all jetted NLS1s clearly show signs of interaction \citep{Varglund22, Vietri22}. The link between jets and host galaxy, therefore, remains unclear. 
   
\subsection{Central engine}
\label{sec:bhmass}
The optical spectrum of NLS1s is used to derive two of their main physical parameters, that is, the black hole mass and the Eddington ratio. In AGN, the black hole mass is usually measured assuming that the gas in the BLR is virialized. If this is the case, then 
\begin{equation}
    M_{BH} = f\frac{R_{\rm BLR}v^2}{G} \; ,
\end{equation}
where R$_{\rm BLR}$ is the radius of the BLR, $v$ is the gas rotational velocity, G is the gravitational constant, and $f$ is a correction factor that depends on the geometry of the BLR. There are various ways to measure the three unknown parameters of this equation. 

The best way to determine the BLR radius is by means of reverberation mapping (RM) campaigns \citep{Peterson93}. This technique is based on the fact that broad lines respond to variations in the continuum of the central engine with a delay due to the light-travel time. In simple terms, long-term spectroscopic monitoring allows us to estimate this delay $\tau_{\rm BLR}$, and to derive the BLR radius simply as R$_{\rm BLR} = c\tau_{\rm BLR}$, where $c$ is the speed of light. This is a responsivity-weighted light-travel radius. Converting it into a physical BLR geometry and virial mass introduces model-dependent systematics. Several RM campaigns have been carried out in the last decades \citep[e.g.,][]{Peterson04, Grier12, Fausnaugh17, Derosa18, Dallabonta20, Bao22, Oknyansky23}, in some cases by using photometric observations instead of spectroscopy \citep{Shablovinskaya23, Ma23}. Although extremely accurate, this method has a major downside, which is the need for long observing campaigns with multiple telescopes, while the interpretation of the lag and the virial factor remains affected by the poorly known BLR geometry. However, the results of RM campaigns have been used to calibrate multiple scaling relations that link the BLR radius with the luminosity of the continuum or, in some cases, of emission lines \citep{Kaspi00, Greene10, Bentz13}. Furthermore, it has been shown that this relation between the luminosity and the radius of the BLR has an additional dependence on the Fe II strength \citep{Du19}. When R4570 is large, the observed time lags, i.e. the BLR radius, are smaller than expected \citep{Grier17, Du18}. This issue is particularly significant in NLS1s, where the R4570 is typically higher than in other AGN classes. 

As for the gas rotational velocity, the typical proxy used to measure it is the FWHM of the H$\beta$ line. Given that the NLS1 classification depends on this quantity, the line is normally present in the optical spectrum. Other options, more rarely used, are the FWHM of H$\alpha$, which is mostly used for the NLS1-like sources found by \textit{JWST} \citep{Dallabonta25}, and that of Mg II $\lambda$ 2798. It has been shown, however, that the second-order moment of the H$\beta$ broad component is a better indicator for the gas velocity \citep{Peterson04, Dallabonta20}, since it is in principle less affected by the geometry of the BLR than the FWHM.  
 
The factor $f$ includes basically every unknown due to the structure and inclination of the BLR, and it is arguably the largest source of uncertainty when estimating the black hole mass. Assuming, for example, a disk-like BLR dominated by Keplerian motion, which has been observed in some sources \citep{Gravity18}, when observed pole-on, there would be no velocity component directed along the line of sight that could be measured in the emission lines. This would cause an underestimate of the gas rotational velocity and, in turn, of the black hole mass \citep{Decarli08}. In a more sphere-like BLR, instead, this effect would be significantly less evident. In the case of NLS1s, the BLR geometry probably falls in the latter case. The Lorentzian profile observed in the permitted lines of NLS1s \citep{Sulentic00, Marziani01, Komossa08, Cracco16, Berton20a, Paliya24} likely indicates the presence of turbulent vertical motion of the clouds, possibly originating in a disk wind, on top of the Keplerian motion around the black hole \citep{Gaskell09, Kollatschny11, Goad12, Kollatschny13a}. The effect of this different geometry is that the $f$ factor does not have a major impact on the mass estimate of NLS1s, which can typically be determined with a precision between 0.4 and 0.5 dex \citep{Grupe04a, Komossa07, Cracco16, Berton21b}. In the literature, there are several attempts to estimate the $f$ factor in AGN, but typical values are between 0.8 and 5 \citep{Mandal21}. 

Finally, the BLR radius and kinematics can also be constrained directly through spectro-interferometry in the brightest nearby AGN \citep{Gravity20}. However, this technique is still relatively limited and has currently only been applied to a handful of NLS1s (e.g., Mrk 1239, \citealp{GRAVITY24}). The newly updated version of the instrument, GRAVITY+, will allow to apply this technique to several new NLS1s at larger distances. 

All the previous considerations allow us to measure the black hole mass of AGN and, in particular, NLS1s. Surveys using single-epoch spectroscopy, often backed up by reverberation mapping, typically show that NLS1s harbor black holes whose mass is between $10^6-10^8$ M$_\odot$ \citep[see][and references therein]{Peterson11, Komossa18}, slightly lower than the values normally observed in broad-line Seyfert 1 (BLS1) galaxies, which are closer to $10^7-10^9$ M$_\odot$ \citep{Paliya24}. Part of this difference is built into the FWHM-based definition; nevertheless, the virial mass also depends on R$_{\rm BLR}$ and on the BLR geometry. On average, non-jetted NLS1s tend to have slightly lower mass than (known) jetted NLS1s \citep{Berton15a}, although this could be a selection effect due to the jet power scaling nonlinearly with the black hole mass \citep{Foschini14}. The discovery of gamma-ray NLS1s (see Sect.~\ref{sec:gamma}) triggered a heated debate on the mass of these objects. As mentioned before, some authors proposed that the narrowness of permitted lines could be due not to the low black hole mass, but to an inclination effect \citep{Decarli08} or to other causes such as a high radiation pressure \citep{Marconi09}. These ideas have been used to reconcile the contradiction between the jet paradigm \citep{Laor00}, which observationally associated powerful relativistic jets with high black hole masses without defining a sharp physical threshold, and the low-mass derived for NLS1s. However, it is now evident that the paradigm was an observational bias that has no root in reality. The host galaxy morphology (Sect.~\ref{sec:host}), reverberation mapping studies \citep{Wang16}, and physical arguments \citep[e.g.,][]{Berton19b, Foschini20, Berton21b}, clearly proved that, regardless of the presence of powerful relativistic jets, NLS1 black holes are undermassive compared to the general AGN population. 

The typical luminosity of NLS1s is comparable to that of regular BLS1s. Usually, their bolometric luminosity is estimated using scaling relations. Typical examples are the linear relation between $\lambda$5100 \AA{}\ continuum and the total luminosity \citep{Kaspi00}, or between the emission lines strength and the bolometric luminosity \citep{Koratkar91}. Accounting for their lower black hole mass, the luminosity yields that NLS1s are high-Eddington sources \citep{Boroson92, Sulentic00, Sulentic02, Grupe04}. The Eddington ratio, defined in Eq.\ref{eq:edd}, is the main driver of several observed properties in NLS1s \citep{Marziani01, Shen14}. A high Eddington ratio, for example, can effectively produce the powerful winds and outflows that are often observed in NLS1s, such as in the C IV or [O III] lines, but also in radio and, possibly, in cold gas. The structure of the accretion flow in these high-Eddington sources is theoretically expected to depart from the standard Shakura-Sunyaev accretion disk \citep{Shakura73} and approach a slim-disk regime \citep{Abramowicz88} with an advection-dominated, but not necessarily radiatively inefficient, inner flow \citep{Marziani14}.
In super-Eddington accretion flows, photon trapping reduces the radiative efficiency of the disk: a significant fraction of the radiation generated in the inner regions is advected inward with the flow rather than escaping. As a result, the emergent luminosity can remain near or below the Eddington limit even when the mass accretion rate is formally super-Eddington. Consequently, estimates of the Eddington ratio based on the observed luminosity may appear sub-Eddington despite a genuinely super-Eddington mass accretion rate \citep{Abramowicz13}. The accretion rate also plays a crucial factor in the different time lags measured via reverberation mapping. As previously mentioned, the sources with the largest lag deviations are also those with the highest Eddington ratio and, in turn, the highest R4570 \citep{Du18}. Finally, it is worth noting that any estimates of the black hole mass based on the assumption that NLS1s harbor a standard disk are inevitably inaccurate, and typically lead to an overestimate of the mass up to two orders of magnitude \citep{Calderone13, Viswanath19}.

\section{X-rays}

\begin{figure*}
    \includegraphics[width=\columnwidth]{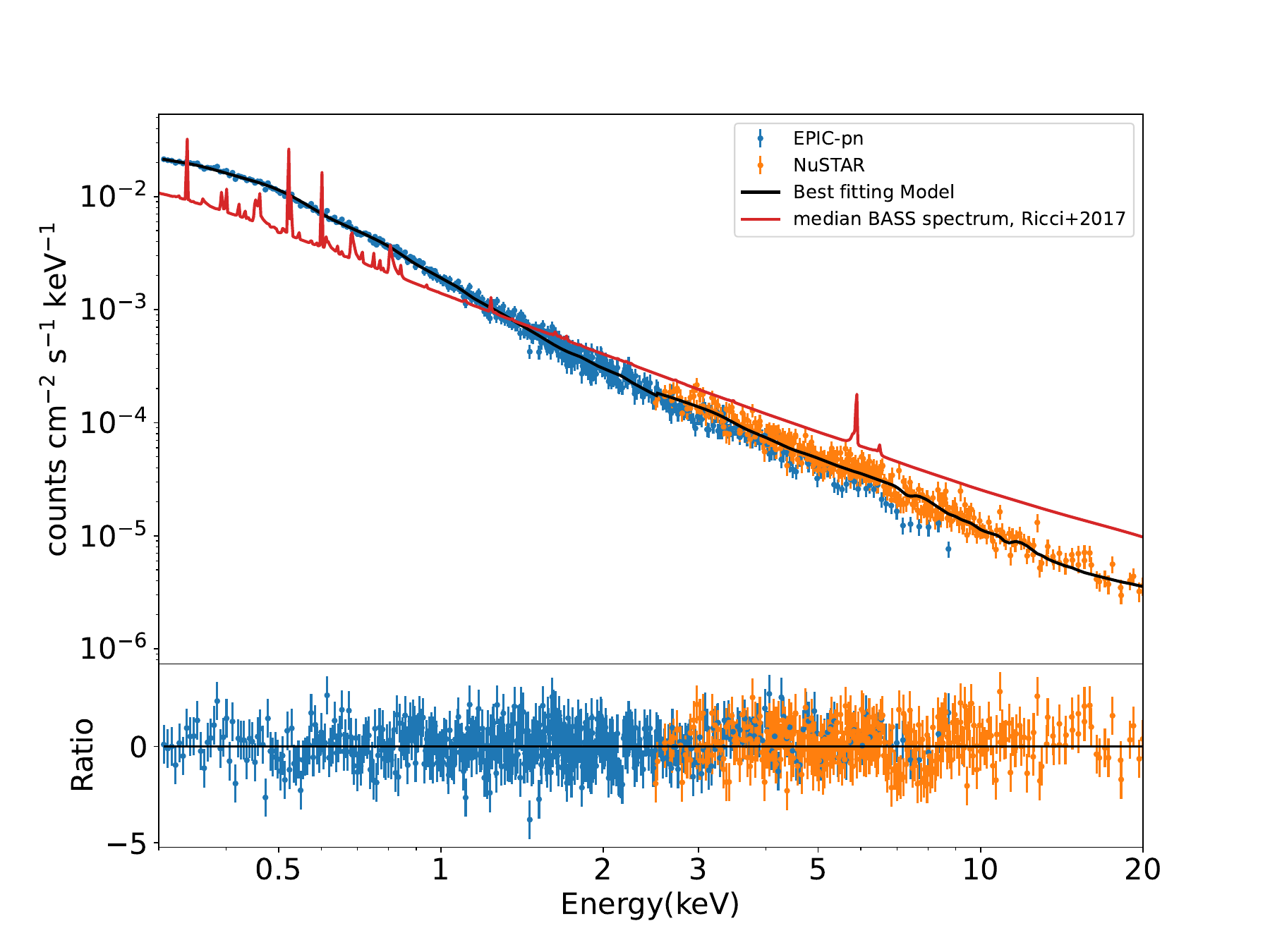}
    \includegraphics[width=\columnwidth]{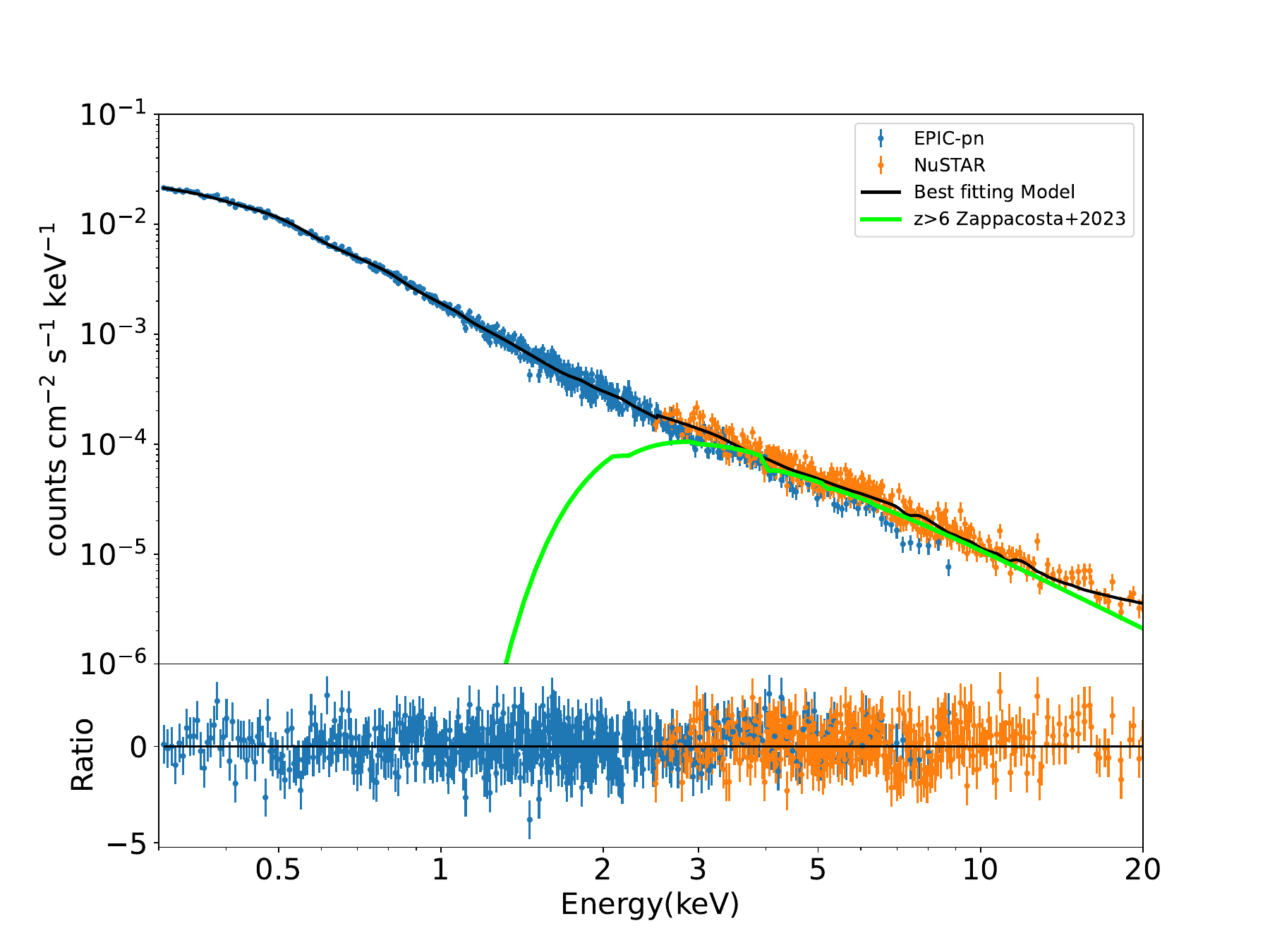}
    \caption{Image of the X-ray spectrum of a typical NLS1 (TON\,S180 \citep{Matzeu20}). We show data, best fitting model and ratio residuals for the EPIC-pn (blue) and \textit{NuSTAR} fpma+fpmb (orange) spectra. The total model (solid line) is decomposed into its main physical components following \citealt{Matzeu20}: thermal emission from the accretion disk (optical--UV), warm optically thick Comptonization producing the soft X-ray excess, hot optically thin Comptonization responsible for the primary X-ray continuum, and relativistic reflection from the accretion disk. The two panels show the same TON S180 spectrum against different comparison samples to avoid overcrowding: the left panel includes the mean spectrum (red) of the local hard X-ray selected AGN in the BAT AGN Spectroscopic Survey (BASS, \url{https://www.bass-survey.com}; $z_{\rm med}<0.035$; \citealt{Ricci17}), while the right panel includes a power-law model with the average photon index ($\Gamma \sim 2.4$) of the HYPERION hyperluminous QSOs at $z \gtrsim 6$ \citep{Zappacosta23,Tortosa24}.}
    \label{fig:Xray_spec}
\end{figure*}

X-ray emission in AGN mainly originates from a hot corona of relativistic electrons, located in the vicinity of the supermassive black holes (SMBHs). Thermal UV/optical photons emitted from the accretion disk are inverse-Compton scattered by the hot electrons into the X-rays \citep[e.g., ][]{Sunyaev80, Haardt93}. The Comptonization leads to a power-law continuum, characterized by a photon index, $\Gamma$, and extending up to energies determined by the plasma electron temperature ($kT_e$), where a cut-off, $E_c$, occurs. The primary X-ray continuum is reprocessed \citep{Matt91} by the cold neutral circumnuclear medium (i.e., the accretion disc or the molecular torus) and gives rise in the X-ray spectrum to a reflection bump at around 30\,keV and a broad iron, K$\alpha$ line emission at around 6.4\,keV. Eventually, the X-ray spectrum of AGN could be characterized by a rise of the spectrum below 1-2\,keV, the so-called soft-excess \citep{Arnaud85,Bianchi09}. The soft-excess is an excess of emission at soft X-ray energies relative to the extrapolation of the primary continuum. Several physical mechanisms have been proposed to explain this component, including (i) warm, optically thick Comptonization in a low-temperature corona \citep[“warm corona” scenario, e.g.][]{Done12,Petrucci18}, (ii) relativistically blurred reflection from the inner accretion disk \citep[e.g.][]{Ross00,Crummy06}, and (iii) emission associated with high accretion rate flows, such as those expected in radiation-pressure dominated or slim-disk regimes \citep[e.g.][]{Abramowicz88,Mineshige00,Wang14}. The relative importance of these mechanisms is still debated and may vary from source to source.

Ionized gas outflows from rapidly accreting SMBHs may play a significant role in regulating both the inflow and outflow of material onto and from the SMBH \citep[e.g.,][]{King03,King15,Fiore17,Cicone18,Laha21}. In the X-ray spectrum, the presence of outflows is typically indicated by absorption features. Low-ionization absorbers, with ionization fraction \logxi $\leq2$, which exhibit relatively slow outflow velocities ($v_{\rm out} \sim 100-1000$\, km s$^{-1}$), are detected in approximately 65\% of soft X-ray spectra ($E\leq 2$ keV) of nearby AGN and are frequently referred to as warm absorbers \citep[WAs, e.g.,][]{Halpern84,Blustin05,Laha14}. Their detectability depends strongly on spectral resolution and signal-to-noise ratio, so a non-detection in low-quality CCD spectra does not necessarily imply their absence \citep{Blustin05,Laha14}. Additionally, more than 30\% of AGN observed in X-rays show evidence of ultra-fast outflows \citep[UFOs, e.g.,][]{Chartas02,Pounds03a,Pounds03b,Braito07,Tombesi10,Tombesi13,Nardini15}. UFOs are highly ionized absorbers, \logxi $\sim3-6$, marked by blueshifted absorption lines from Fe XXV and XXVI, and typically exhibit outflow velocities of $v_{\rm out} \sim 0.1$ c, with some reaching near-relativistic speeds up to $v_{\rm out} \sim 0.5$c  \citep[e.g., ][]{Reeves18,Luminari21}. UFOs are often observed in the X-ray spectra of NLS1s. These fast outflows are believed to be driven by the intense radiation pressure from the accretion disk, which is especially pronounced in the case of NLS1s due to their smaller BH masses and higher Eddington ratios \citep[e.g.,][]{Leighly97,Fabian09,Kara17,Krongold21,Tortosa22}. Indeed, one of the mechanisms invoked as a key ingredient at all cosmic epochs for the launching of powerful outflows is the super-Eddington accretion \citep[e.g.,][]{Jin17a}.

In the X-ray band, NLS1s show a rapid and large variability, with the break time scale, defined as the inverse of the break frequency at which the power spectral density (PSD) slope changes from a flatter to a steeper power-law, having the behavior observed in most AGN, increasing proportionally with black hole mass and decreasing with increasing accretion rate \citep{Uttley05,McHardy04,McHardy06,Ponti12,Tortosa23b}. 
NLS1s show complex spectral properties such as a steep X-ray spectral slope ($\Gamma \sim 2.0 - 2.2$, e.g. \citealt{Brandt97,Vaughan99}, see left panel of Fig.\,\ref{fig:Xray_spec}) possibly driven by enhanced EUV/soft X-ray emission \citep{Boller96,Brandt97,Leighly99a,Leighly99b} and evidence for cold and ionized absorption, partial covering, and strong features of reprocessed radiation, as well as a soft excess below 1 keV and a dip at $\sim$7 keV \citep{Wang96, Komossa00,Fabian02,Crummy06}. The EUV/soft X-ray emission enhancement is commonly interpreted as a consequence of the high Eddington ratios typically observed in NLS1s. For a given luminosity, their lower black hole masses imply higher accretion disk temperatures, shifting the thermal emission toward the EUV/soft X-ray band. In addition, high accretion rates may produce a strong warm, optically thick Comptonizing region (``warm corona'') that efficiently up-scatters disk photons into the soft X-ray range. Radiation-pressure dominated or slim-disk structures expected at high accretion rates may further modify the inner disk emission and contribute to enhanced EUV/soft X-ray output \citep{Pounds95,Mineshige00,Done12,Petrucci18}.

\citet{Pounds95}, studying the spectrum of the NLS1 RE\,J1034+39, suggested that the steep spectrum of these objects could be the result of the much stronger Compton cooling of the corona by the strong radiation field from the super-Eddington disk, which may cause also a lower temperature of the AGN corona, as seen also by \citet{Kara17,Tortosa22,Tortosa23}. Indeed, optically-thick geometrically-thin accretion discs can explain several key features of the observed spectral energy distribution (SED) of AGN with moderate Eddington ratios ($\lambda_{\rm Edd}=L_{\rm bol}/L_{\rm Edd}\in[0.01;0.3]$, \citealt{Koratkar99, Capellupo15}). At higher accretion rates, the disc is expected to become geometrically thick (i.e., slim disc, \citealt{Czerny19}), and the nature of the accretion flow is expected to change dramatically by photon trapping through electron scattering in the dense matter and advection cooling. Moreover, strong gas outflows are naturally expected during super-Eddington accretion episodes \citep{Ballantyne11,Zubovas12,Jiang24} due to the intense radiation pressure associated with these events. The presence of outflowing disc winds has also been observed in some high-redshift quasars (QSOs) accreting close to the Eddington limit \citep{Chartas03, Lanzuisi12, Vignali15, Lanzuisi16} and in ultraluminous X-ray sources (ULXs) \citep{Pinto17}. 

While the X-ray emission of most NLS1s is dominated by the disk corona, the class of jetted NLS1s, and especially those emitting gamma-rays, is significantly different. Indeed, beamed jetted NLS1s show a bimodal distribution in their photon index, with approximately half having a hard spectrum (i.e. $\Gamma <2$), similar to regular Seyfert 1 galaxies, while the others show a softer spectrum ($\Gamma\sim2.7$), more typical of non jetted NLS1s \citep{Foschini15}. This clearly indicates that the jet itself strongly contributes to the X-rays, as observed in most jetted AGN \citep{Leighly99a, Leighly99b, Piconcelli05}. In particular, the spectra of gamma-NLS1s are typically dominated by the jet emission from 0.3 to 70 keV, with some additional components visible from time to time \citep{Berton19b, Yao19, Yao23}. A particularly interesting case is that of 1H 0323+342, the only gamma-NLS1 that to date showed a (weak) Fe K$\alpha$ emission line \citep{Abdo09c, Kynoch19, Paliya19}, although it was not present in all observations \citep{Landt17, Berton19b, Dammando20}. This source was detected in hard X-rays already with early observations with INTEGRAL \citep{Bird07, Malizia07}, but only later its spectral variability revealed the presence of a relativistic jet \citep{Foschini09}. Recently, the long-term analysis of Swift data revealed that this object transitions between three different regions in the photon index-flux plane. This result can be interpreted in the framework of disk instability \citep{Czerny09} and jet activity, with the source changing from a jet-dominated state to a disk-dominated one \citep{Rosa25}.

\section{Gamma-rays}
\label{sec:gamma}
\subsection{Gamma-ray emitting NLS1s}
With very few exceptions \citep{Abdo10, Angioni19, Abdollahi20}, the origin of gamma-rays in extragalactic sources lies in a beamed relativistic jet, i.e., closely aligned with the line of sight. This normally occurs in the AGN class of blazars, which are the brightest non-transient sources in the Universe \citep[e.g.,][]{Madejski16, Hovatta19, Blandford19}. The SED of blazars has a very characteristic shape, as it is clearly double-peaked. The low-energy peak corresponds to synchrotron emission produced by relativistic electrons accelerated in the magnetic field of the jet. The high-energy peak originates instead via the inverse Compton mechanism. The electrons of the jet are cooling down while at the same time up-scattering at high energy either the UV/optical/NIR photons coming from the circumnuclear environment, e.g. the accretion disk, the BLR or the torus, or the synchrotron photons that are produced by the same population of electrons. These two mechanisms are known as external Compton (EC) or synchrotron self-Compton (SSC), and they are normally found in the two main classes of blazars, flat-spectrum radio quasars (FSRQs) and BL Lacertae objects (BL Lacs), respectively. The behavior of blazars is summarized by the so-called blazar sequence, which provides the physical interpretation we described for the observed properties of these sources \citep{Fossati98, Ghisellini98}. In general, high-power sources such as FSRQs have both peaks at relatively low frequency (synchrotron peak in radio, EC peak at MeV/GeV), while low-power sources have both peaks at high frequency (synchrotron peak in X-rays, SSC peak at TeV, \citealp{Ghisellini16, Prandini22}). The blazar sequence fits very well in a scenario where sources with a high jet power are characterized by a high-density circumnuclear environment where electrons can cool down efficiently, while low-power sources live in a low-density environment where electron cooling is significantly less effective \citep{Heckman14}. 

This simple view of blazars held until the launch of the \textit{Fermi} satellite in 2008. Soon after this milestone event for high-energy astrophysics, a handful of unexpected guests showed up at the party. Four NLS1s, 1H 0323+342, PMN J0948+0022, PKS 1502+036, and PKS 2004-447, were detected at GeV energies \citep[][see the reviews by \citealp{Foschini11, Foschini12}]{Abdo09a, Abdo09b, Abdo09c}. One of them, PMN J0948+0022, underwent several gamma-ray flares in the following years, reaching an observed luminosity up to 10$^{48}$ \ergs \citep{Foschini11c} and a gamma-ray variability of the order of minutes \citep{Itoh13}. This indicates that these objects harbor powerful relativistic jets aligned with our line of sight, exactly as in the two classic blazar classes, and, therefore, they constitute a third class of blazars. 

A few comprehensive reviews have already been written on this class of sources \citep{Foschini11, Foschini12, Komossa18, Paliya19, Dammando19, Foschini20}, but here we will summarize a few essential points. The SEDs of gamma-NLS1s share similar peak positions with FSRQs, but have comparable jet power to BL Lacs \citep[][see also Fig.~5 of \citealp{Paliya13}]{Foschini15}. This behavior cannot be explained simply in terms of electron cooling as predicted in the blazar sequence. An additional ingredient is necessary, i.e. the black hole mass. As discussed in Sect.~\ref{sec:bhmass}, NLS1s are powered by undermassive black holes relative to those in FSRQs and in the general broad-line AGN population, accreting close to the Eddington limit. These gamma-ray NLS1s are not different. Since the jet power scales nonlinearly with the black hole mass \citep{Heinz03, Foschini14}, an undermassive black hole naturally explains the lower power observed in NLS1 jets \citep{Foschini17}. Indeed, it was confirmed that gamma-NLS1s represent the low-luminosity tail of the radio luminosity function of FSRQs, as would be expected if these objects were low-mass and low-power counterparts of the most powerful blazars \citep{Berton16c}. As a class, gamma-NLS1s seem to be on the high-mass tail of the NLS1 black hole mass distribution \citep{Foschini15}, thus connecting directly with FSRQs, but this could be a selection effect due to the higher jet power and enhanced detectability of high-mass objects. It is possible that low-mass gamma-NLS1s exist and that they will be detected in the future in case of flaring activity. After the discovery of the first four gamma-NLS1s, \textit{Fermi} kept detecting known NLS1s or, vice versa, some new gamma-ray sources were identified as NLS1s \citep{Dammando12, Liao15, Yao15, Dammando16, Berton17, Yang18, Romano18, Paliya18, Lahteenmaki18, Yao19, Paiano19, Rakshit21b, Li23, Gabanyi25}. Furthermore, a detailed analysis of the most recent catalog of extragalactic radio sources revealed that several NLS1s could be misclassified as FSRQs or just grouped with the large category of unidentified \textit{Fermi} objects \citep{Foschini21, Foschini22}. The existence of gamma-ray emitting NLS1s, in conclusion, proves that the blazar sequence is valid only if the sources have comparable black hole mass, but it needs to be updated if different mass ranges are included. 

\subsection{The parent population}
Basically, all gamma-ray NLS1s are bright radio sources showing a flat spectrum \citep{Yuan08, Berton18a, Shao25}. However, there are two exceptions to this rule: 3C 286 and PKS 2004-447 \citep{Berton17, Berton21b}. Both of them are part of the Fermi catalog \citep{Abdollahi20}, but unexpectedly show a steep radio spectrum. In particular, they belong to the class of compact steep-spectrum (CSS) radio sources \citep{Peacock82, Oshlack01, Schulz15}. CSS sources are one of the many classes of AGN that can be labeled as peaked sources (PS), i.e. objects characterized by a peak in their radio spectrum above $\sim$100 MHz. In general, PSs harbor relativistic jets confined within their host galaxy. There are several mechanisms to explain the small size of PSs, like interaction with a dense environment or intermittent jet activity, but a widely accepted hypothesis is that most of them are genuinely kinematically young sources, that is, the progenitors of radio galaxies \citep{Fanti95, Odea98, Orienti16, Odea21}. As we will show later, NLS1s are also considered as the progenitors of other AGN classes, such as BLS1s \citep{Mathur00}. For this reason, many authors suggested a potential connection between PSs and NLS1s \citep{Oshlack01, Komossa06, Gallo06, Yuan08, Caccianiga14, Gu15, Schulz15, Caccianiga17, Liao20, Berton21c, Jarvela22}. In particular, it was hypothesized that some CSSs could represent part of the parent population of gamma-ray emitting NLS1s, i.e. jetted NLS1s observed at a large angle. A detailed analysis of the radio luminosity function of jetted NLS1s with a flat spectrum (beamed) revealed that the specific class of PSs known as low-luminosity compact (LLC) objects could potentially be the same kind of sources only seen at larger angles \citep{Berton16c}. More recently, it was shown that the host galaxy of LLCs is also typically a late-type galaxy, as normally observed in NLS1s \citep{Vietri24}, thus strengthening the hypothesis of LLCs as parent sources. 

Despite this interesting result, the puzzle of the parent population remains wide open. Even if LLCs could be part of it, the current number of known gamma-ray emitting NLS1s is between 20 and $\sim$40. Simple geometrical arguments show that for $n$ beamed objects there must be at least $\sim$2$n\Gamma^2$ parent population sources, where $\Gamma$ is the bulk Lorentz factor of the relativistic jets. Assuming a typical value of $\Gamma \sim 10$ \citep{Abdo09c}, the number of parent sources could go up to 16,000. The number of known LLCs is significantly lower than this. Even halving down the bulk Lorentz factor and considering the most conservative number of gamma-NLS1s, LLCs only cannot be the whole picture. Therefore, the majority of the parent population is still evidently unknown, with the extreme variability found in NLS1s possibly playing a role in this \citep{Lahteenmaki18, Berton20a, Jarvela24}. It is very interesting to note that, because of the discovery of a blazar at the epoch of reionization, the exact same problem is currently open for jetted AGN in the early Universe \citep{Banados25}.

\section{The current view of NLS1s}
From the panchromatic view that we presented, the safest conclusion we can draw is that NLS1s have low black hole mass, high Eddington ratio, and are usually hosted in spiral galaxies with a pseudobulge. Many of their observational features are clearly driven by their mass and Eddington ratio, and little to no doubt remains on this specific point of their nature. This statement is also true for jetted NLS1s. Despite some debate, there is still little to no proof that the basic physical properties of jetted NLS1s are different from those of non-jetted sources. The most intriguing aspect, however, is why they have a low mass and a high Eddington ratio: what is the root cause of NLS1s' properties? 

Back in 2000, Smita Mathur suggested that, because of their peculiar properties, NLS1s could represent an early stage in AGN life cycle or, in other words, they could be young AGN \citep{Mathur00}. The word "young" is used here not to refer to the actual age of the central black hole. It is most likely that all seeds of supermassive black holes originated in the early Universe and are therefore the same age. The point we want to make is that NLS1s' black holes did not grow as much as in other sources, such as BLS1s, and only in recent times have they started to efficiently accrete and grow. This scenario agrees with the large-scale environment in which NLS1s are typically found, which usually shows a low density \citep{Jarvela17, Jarvela17a}. It is possible that, with such an isolated neighborhood, NLS1s mostly accrete via secular processes, which requires a long time before it can trigger nuclear activity \citep{Orbandexivry11}. If this is the case, NLS1s reside in galaxies that were quiescent for most of their life and only recently became active. A seminal case could be that of Mrk 1044. If the circumnuclear star-forming ring located 500 pc away from its nucleus has not been touched by the AGN ionizing continuum, this puts an upper limit to its age to $\sim$1600 years \citep{Winkel22, Winkel23}. Other objects studied by the Close AGN Reference Survey revealed the presence of a correlation between the size of the extended NLR in nearby AGN and the black hole mass \citep{Husemann22}, and the lack of significant feedback in these sources, possibly due to the fact that the galaxy nucleus has not been active for a long time and hence had no time to impact its host galaxy yet \citep{Smirnova22}. This scenario is consistent with the idea that the quasar main sequence \citep{Sulentic15} represents a sort of "arrow of time" of black hole growth \citep{Fraixburnet17a, Fraixburnet17b}. Within this interpretation, the decrease of EW([O III]) toward large R4570 is consistent with a high-Eddington phase in which the NLR is less prominent, although it is not by itself a unique evolutionary clock \citep{Boroson92,Shen14}. Young sources would be those with a lower black hole mass, a high Eddington ratio, and a high R4570, i.e. abundant iron emission. A key element to tell the evolutionary stage of an AGN could be the broad line profile, with Lorentzian sources on the young-end of the sequence, where the BLR is still dominated by the turbulent gas reaching the nucleus (see \citealp{Berton20a}, Sect. 5.4). 
 
The origin of iron in NLS1s is unclear, but these atoms form in supernovae. The enhanced star formation rate of NLS1s could lead to the explosion of several massive stars, thus increasing the metallicity of the gas \citep{Chen09, Million11}. The typical velocity of supernovae ejecta is $\sim$5000 \kms, therefore the iron can cross a star forming ring as large as that in Mrk 1044 in just 10$^4$ years, which is much less than the typical timescale of an NLS1 event (up to 10$^8$ years, \citealp[]{Komossa18}), and comparable to the duration of a single activity cycle due to disk instabilities \citep{Czerny09}. The high Eddington ratio, finally, could be responsible for the outflows observed in the emission lines by means of radiation pressure, but their physical properties are still largely undetermined \citep{Rakshit18a}. 

This scenario is not dissimilar from what was proposed in 2000. Two of the main observational tests proposed by Mathur to confirm this model were investigating the emission line properties of high-redshift quasars and studying their X-ray spectra. Back then, all of this was still unknown, but today we are finally starting to progress on these points as well. 

\section{High-$z$ vs low-$z$}

In this section, we will examine the potential relation between high-$z$ AGN and NLS1s. We note that a comprehensive review of the multiple results already produced by \textit{JWST} since its launch is beyond the scope of this work. Here, we discuss specific points which help to highlight the importance that NLS1 studies can have in furthering our understanding of the properties of high-$z$ sources. In the comparison below, we distinguish direct observables, such as line widths, line profiles, continuum shapes, and detection fractions, from inferred quantities such as black hole mass, bolometric luminosity, and Eddington ratio. The latter are generally obtained using single-epoch virial relations and bolometric corrections calibrated on local AGN; their application to \textit{JWST}-discovered broad-line AGN and LRDs is therefore model dependent, particularly if the BLR geometry, covering factor, line-broadening mechanism, or host contribution differs from the local calibration samples.

\subsection{Restframe optical properties and black hole mass}

Luminous QSOs, with bolometric luminosity $>10^{46}$ \ergs at $z$ $\geq$6, were firstly discovered in the early 2000s \citep{Fan00}. Hereafter, ``high-$z$ quasars'' denotes this luminous population, whereas ``\textit{JWST}-discovered broad-line AGN'' and LRDs refer to the generally fainter sources discussed below. Currently, there are more than 500 known quasars at such redshift (for a recent review, see \citealp{Fan23}). They already show very large black hole masses, above $\sim10^8$ M$_\odot$, with very high accretion rate. For instance, the sample of high-$z$ quasars described by \citet{Mazzucchelli23} has a median Eddington ratio of 0.76. It is important to notice that these estimates are based on the Mg II scaling relations, but that more recent estimates based on H$\alpha$ yield consistent results \citep{Yang23}. High accretion rate is a property shared with low-$z$ NLS1s, but their black hole mass is radically different, making high-$z$ quasars more similar to more evolved low-$z$ sources, such as BLS1.

However, studies have shown that the fainter spectroscopically confirmed high-$z$ AGN discovered by \textit{JWST} exhibit relatively narrow ‘broad’ permitted lines, placing them within the formal definition of NLS1s \citep[e.g.,][]{Larson23}. It is worth noting, though, that the classification of NLS1s is based exclusively on the H$\beta$ line, which is only rarely visible in these high-$z$ sources. The H$\alpha$ line is more commonly used to study AGN in these objects. Its FWHM is obviously related to that of H$\beta$, but not in a 1:1 relation. In large NLS1s samples it was found that FWHM(H$\alpha$) $\sim$ 0.8-0.9 FWHM(H$\beta$). The smaller H$\alpha$ width is generally attributed to radial stratification and optical-depth effects in the BLR, with H$\alpha$ preferentially emitted at somewhat larger radii. Therefore the formal threshold for NLS1s when using H$\alpha$ as a reference should be moved to 1600-1800 \kms\ \citep{Rakshit17a, Paliya24}. It is also worth reminding that it is hard to distinguish between the AGN types whose optical spectra resemble each other, such as NLS1s and BLS1s, intermediate-type AGN, and intermediate mass black holes, unless the spectra have a very high signal-to-noise ratio ($\gtrsim$ 30). This should be kept in mind when interpreting \textit{JWST} data. 

Using these thresholds, more than half of the \textit{JWST}-identified AGN at high redshift should be classified as NLS1s (see, for example, Figure~11 in \citealp[]{Maiolino25} for a broad H$\alpha$ FWHM distribution). In addition to the narrow BLR Balmer lines, these high-$z$ AGN also share other properties with NLS1s \citep{Larson23, Harikane23, Kocevski23, Maiolino24b, Napolitano25}. Their virial black hole mass estimates, derived from their H$\alpha$ broad component, fall in the range of 10$^5$-$10^8 M_\odot$. They exhibit consistently high Eddington ratios ($\gtrsim$0.1), several of them accreting at super-Eddington rates. These properties point to these high-$z$ AGN currently going through an evolutionary phase whose characteristics are similar to those of the local Universe NLS1 population. An important difference is the lower metallicity inferred for the host-galaxy/ISM gas of many high-$z$ AGN; this should not be automatically extended to the BLR, since luminous high-$z$ quasars can already show chemically mature broad-line gas \citep{Mazzolari25, Scholtz25}. 


Little Red Dots (LRDs) further complicate this view. LRDs are compact objects found at z $\geq$ 4 with very red near-infrared colors, V-shaped SED in the optical/NIR bands, and they have been identified in multiple \textit{JWST} observations \citep[e.g.,][]{Labbe23, Kocevski23, Harikane23, Greene24, Matthee24}. The debate on the nature of these objects is still ongoing, since their origin could be due to both intense star formation or AGN activity \citep{Barro24, PerezGonzalez24}. Although we do not observe a similar V-shaped SED in NLS1s (but note that the corresponding optical/NIR SED region remains relatively unexplored in NLS1s), what they seem to have in common is that they exhibit broad hydrogen lines with FWHM of less than $\sim$2000 \kms, comparable to NLS1s. These narrow lines indicate that their black hole mass could be rather low, between 10$^5$-10$^7$ M$_\odot$, and characterized by super-Eddington accretion \citep{Maiolino24a, Maiolino24b}. These mass and accretion estimates are inferred through local single-epoch virial calibrations and bolometric corrections and should therefore be regarded as uncertain for LRDs. It is worth noting that, according to \citet{Matthee24}, there is a significant decline in the number of sources with $M < \sim10^7$ $M_\odot$ among LRDs. This is likely due to a selection effect, since at fixed mass, only the highest accretors can be detected. \citet{Matthee24} also suggest a rapid evolution of these sources from star formation dominated, into young AGN with an NLS1-like profile, and then into an adult AGN with a very broad line and similar to BLS1s, showing also inflows/outflows (their Figure 20). Unlike local AGN, where the nucleus is mostly unobscured, recent studies propose that these sources are completely engulfed in a dense cocoon of gas \citep{Naidu26, Degraaff25, Inayoshi25, Inayoshi25a}, which well reproduces their Balmer breaks and absorption features. Furthermore, it could explain their observed radio weakness, which would be due to free-free absorption \citep{Mazzolari25, Gloudemans25, Latif25}. This form of chaotic accretion could be compared to the vertical structure of the BLR that is found in NLS1s, where the young age also leads to a turbulent motion of the clouds \citep{Goad12, Kollatschny11, Kollatschny13a}, although with less extreme density.

It is worth noting, however, that the line profile observed in most of the LRDs does not match what is seen in NLS1s. The spectral line profile of H$\alpha$ shown, for example, by \citet{Matthee24} is indeed closer to what is seen in intermediate-type Seyferts \citep{Barquin24}, with only a handful of sources exhibiting the typical Lorentzian profile of NLS1s. Balmer absorption features on top of broad emission lines can also be seen \citep{Lin24, Deugenio26}, somewhat similar to a P-Cygni profile \citep{Castor79}, and this is very rare in low-$z$ sources \citep[e.g.,][]{Hutchings02, Ji25}. Related to this, another potential difference is the origin of line broadening in LRDs versus NLS1s. In the latter, broad lines originate due to the gas rotational velocity or by its velocity dispersion around the SMBH, but this mechanism is not the only way to produce broad lines. Early studies on type 1 AGN actually suggested that the origin of the profiles was electron scattering within photoionized gas \citep{Weymann70}. However, the electron scattering optical depth within the typical BLR gas is $\tau_e << 1$, which is too low to explain broad line profiles in local AGN \citep{Davidson79}. In LRDs, though, recent studies revealed that the line profile could be exponential, thus pointing toward an electron scattering origin for the line broadening \citep[][but see the discussion by \citealp{Juodzbalis26} for a different interpretation]{Rusakov26}. If this were true, the black hole mass estimates derived from the emission lines would have to be corrected for this factor, since the virial theorem cannot be applied in these or any other objects when the line broadening is due to something else rotational velocity \citep[e.g., outflows, see][]{King24}. Indeed, by using a theoretical model of a thin and slim accretion disk, \citet{Lupi24} verified that in many cases, for \textit{JWST}-detected sources, a super-Eddington accreting SMBH is a better fit to the data than a standard accretion disk. This suggests BH mass values lower than what is typically derived by up to one order of magnitude, and well within the typical mass and accretion range of NLS1s.

Recently, efforts have been made to detect low-$z$ counterparts of LRDs, by searching for targets with matching properties, such as morphological compactness, V-shaped SEDs, and colors \citep[e.g.,][]{Lin25a, Ma26}. In particular, \citet{Lin26} found three candidates at $z$ = 0.1-0.2 from SDSS. These sources display strong [Fe II] emission lines, that is proposed to originate at intermediate scales between the BLR and the NLR. Based on the properties of these sources, \citet{Lin26} proposed a conceptual model for LRDs' structure, which includes the scenario of a dust-poor gas cocoon that emits thermal radiation that we mentioned earlier. The presence of permitted Fe II is one of the distinctive properties of NLS1s, but most faint \textit{JWST} broad-line AGN show substantially weaker optical Fe II than local objects at the same position along EV1 \citep{Trefoloni25}. Forbidden iron emission is not very common. The only examples are [Fe VII] and [Fe X], which are especially strong in Fe II-poor sources \citep{Pogge11}.

\subsection{X-ray emission}
The X-ray spectral properties of NLS1s resemble those of the QSOs at the Epoch of Reionization (EoR, $z \gtrsim 6$; \citealp{Vito19,Wang21,Zappacosta23,Tortosa24}), although scaled down in mass and luminosity by at least two orders of magnitude. Indeed, as NLS1s typically show a steep soft X-ray photon index, also high-$z$ QSOs show a steep hard X-ray spectrum compared to typical type 1 AGN (see right panel of Fig.\,\ref{fig:Xray_spec}). Unfortunately, due to instrumental limitations and redshift effects, soft excess is often missing or weaker in high-$z$ QSOs' observed data, but the presence of steep soft X-ray photon index often is an indication of a soft excess component, as in NLS1s galaxies. Moreover, NLS1s often have relatively X-ray weak emission compared to UV (lower X-ray to optical-UV ratio, $\alpha_{\rm OX}$). High-$z$ QSOs also tend to show low X-ray emission for their UV luminosity, especially at high luminosities \citep{Martocchia17}. The X-ray weakness is more precisely quantified through $\Delta\alpha_{\rm OX}$, the difference between the measured $\alpha_{\rm OX}$ and the value expected from the $L_{\rm UV}$--$\alpha_{\rm OX}$ relation; a negative $\Delta\alpha_{\rm OX}$ denotes an X-ray deficit relative to quasars of similar UV luminosity \citep{Lusso16}. These similarities in the X-ray emission properties of both NLS1s and high-$z$ QSOs suggest that the high accretion rate plays a central role in shaping the X-ray emission. 

However, recent works \citep[e.g.,][]{Mazzolari25,Maiolino25}, found many \textit{JWST}-selected AGN at high redshift being X-ray weak, with luminosity upper limits ($\rm L_{2-10keV}\lesssim few\times10^{41}$erg/s), which could make them not NLS1-like. For instance, \citealt{Maiolino25}, considering a large sample of broad-line and narrow-line AGN spectroscopically selected in the JADES survey over the \textit{Chandra} Deep Field South \citep[CDFS,][]{Liu17} with 7\,Ms of \textit{Chandra} observations, investigated different possible origins of this X-ray weakness. On one hand, they found an enhanced equivalent width of the H$\alpha$ broad line in these sources, suggesting the BLR to be distributed with a large covering factor around the central source acting as a dense dust-free absorbing medium. The latter can potentially reach Compton-thick column densities, indicating that the BLR of high-$z$ AGN has a higher covering factor compared to what is seen in NLS1s. On the other hand, the authors also find that many of the AGN newly discovered by \textit{JWST} have features in common with the population of NLS1, which are known to have a steep X-ray spectrum (and therefore showing a softer X-ray emission) and/or have moderate-to-high inferred accretion rates. The sample discussed by \citet{Maiolino25} has a mean $\lambda_{\rm Edd}\sim0.15$, so this statement does not imply that the population is generally super-Eddington. These conditions are expected to result into a weaker X-ray emission, especially at high redshift, where higher rest frame energies are probed with \textit{Chandra}.

Another hypothesis that could explain the X-ray weakness of these sources is related to the suppression of the X-ray corona by the radiation pressure and disk winds in high accreting sources, as suggested also for some NLS1s \citep[e.g.][]{Gallo18a}. Strong disk winds, driven by radiation pressure, can interact with the corona, potentially leading to compression, dissipation and shielding effects. The winds may compress the corona, increasing its density and leading to enhanced cooling, which can reduce X-ray emission. In extreme cases, the corona might be dissipated or significantly altered \citep{Gallo18, Yang18a, Ricci20}. Also, dense winds can shield parts of the corona from the accretion disk's radiation, affecting the heating processes necessary for X-ray production \citep{Gronkiewicz23}. For luminous high-$z$ quasars, the rapid BH growth and strong UV radiation imply compact accretion disks and high radiation pressure launching powerful outflows and winds. This could lead to disruption or compression of the X-ray corona, extreme cooling of the corona and shielding effects, resulting in the detected X-ray weakness. An observational proof of this hypothesis could be found in the recently discovered relationship between the X-ray photon index and the velocities of the rest-frame UV disk winds, traced by broad C IV emission lines, in the HYPERION sample of hyperluminous QSOs ($L_{bol}>10^{47}$\,erg\,s$^{-1}$) at $z>6$ \citep{Zappacosta23}. According to this relation \citep{Tortosa24} the steeper the X-ray spectrum, the faster the $\rm C\,\textsc{iv}$ winds, suggesting a physical link between the disc--corona system and the terminal velocities of the accretion disk winds. In a model composed of inner corona--outer accretion disk (see Fig. 5 in \citealt{Tortosa24}), as a consequence of high accretion flows, the inner part of the accretion disk puffs up (i.e. slim disk, \citealt{Czerny09}), due to the high radiation-pressure of UV photons. This plays a role in producing steeper X-rays (as a consequence of coronal cooling induced by the high UV flux from this region) and in launching fast ionized accretion disk winds from inner regions, which are shielded and hence not over-ionized by the central X-ray source flux. This interpretation suggests that X-ray surveys may underestimate the number of high-$z$ AGN due to selection bias against X-ray weak sources, or it hints at a possible evolutionary phase where AGN are X-ray suppressed during early growth \citep{Nanni17, Vito18}, possibly like local changing-look AGN \citep{Ricci17}.

\subsection{Host galaxy and environment}


Sub-mm observations, mainly from ALMA and NOEMA, revealed that in terms of host galaxy properties, high-$z$ quasars are hosted in gas- and dust-rich galaxies, with very high star formation rate ($>$100 M$_\odot$ yr$^{-1}$, \citealp{Decarli18}). They exhibit a variety of morphological and kinematical features, equally distributed among rotation-dominated disks, merging systems, and perturbed morphologies \citep[e.g.,][]{Neeleman21}. The presence of outflows in C+ emission lines is still controversial \citep{Bischetti19, Novak20}. Recent studies report direct evidence of fast molecular outflows via observations of the OH 119 $\mu$m emission line in 70\% of the inspected quasars \citep{Spilker25}. In the last few years, \textit{JWST} opened a window to explore the host galaxy of high-$z$ quasars also in optical restframe. The study of their starlight suggested stellar masses around or above 10$^{10}$ M$_\odot$ \citep{Ding23, Yue24}, and possibly even post-starburst host galaxies \citep{Onoue25}, although caution should be taken given the difficulty in subtracting the contribution of the nuclear emission \citep{Berger26}. This is somewhat reminiscent of what is seen in NLS1s at low-$z$. They are hosted by gas-rich spirals with enhanced star formation, and also show molecular outflows (see Sect. 3.2). However, it is worth noting that the literature on this aspect of NLS1s is still relatively limited, and new studies are necessary to have a more detailed comparison between the two classes. For LRDs, instead, the host morphology seen by \textit{JWST} is very compact, with effective radii smaller than 1 kpc, and they appear relatively dust-free \citep{Juodzbalis26, Setton25}. 

It is interesting to compare the well-known relation found at low redshift between the black hole mass, the stellar velocity dispersion, and the host galaxy stellar mass \citep{Ferrarese00, Kormendy13}. One of the key distinctions between NLS1s and high-$z$ sources, especially LRDs, is their positioning with respect to this relation, that is, the relative mass of their SMBHs compared to their host. Some studies indicate that early SMBHs tend to be overmassive relative to their host galaxy’s stellar mass \citep{Pacucci23, Bogdan24, Juodzbalis24b}, suggesting an early phase of rapid SMBH growth dominating over host galaxy development. This result seems to be confirmed not only with \textit{JWST} observations \citep{Marshall23, Stone23, Stone24, Yue24}, but also with ALMA data \citep{Willott17, Izumi19}. This is consistent with the hypothesis that direct collapse black holes and/or super-Eddington accretion events play a crucial role in SMBH formation \citep{Volonteri21}. However, this may not be true in all cases, and these results should be taken with a grain of salt, given the complexity of deriving stellar masses at high-$z$, and the model dependence of single-epoch black hole masses through the assumed BLR geometry, virial factor, and line-broadening mechanism, and the different evolutionary stage of the sources used at low-$z$ to derive scaling relations. 
In NLS1s, instead, the host galaxy appears to be overmassive compared to the SMBH \citep{Molina24}, or in general not following the normal scaling relations \citep{Mathur01, Bian04, Grupe04, Botte04}. This is suggestive of a later stage in the co-evolutionary process where galaxy growth has caught up or exceeded the rate of SMBH growth. However, this difference does not come as a surprise, given the different cosmological epoch in which NLS1s live. While NLS1s likely form via secular evolution in mostly isolated spiral galaxies, the formation channel of high-$z$ quasars is completely different, and still largely unexplored. 

At $z>$6, although sub-mm observations show the presence of gas-rich companion galaxies in several quasar fields \citep[e.g.,][]{Decarli17}, restframe UV-based observations on Mpc scales suggest a more complex scenario, with a large diversity observed in different fields \citep[e.g.,][]{Ota18, Champagne25}, and several observational limitations and biases \citep[e.g.,][]{Lambert24}. The \textit{JWST} found that high-$z$ luminous quasars are hosted in massive dark matter halos above 10$^{12}$ M$_\odot$, consistent with observations of quasars at cosmic Noon \citep{Eilers24}. Moreover, in the last few years, the number of jetted sources at high-$z$ has strongly increased \citep{Gloudemans22, Endsley22, Ighina25, Belladitta25, Banados25}. Studies of the properties of their host galaxy and environment are still limited, but observations of a few sources showed a potentially larger merger fraction in jetted than non-jetted sources \citep{Khusanova22}, as well as an absence of massive companions \citep{Mazzucchelli25}. In comparison, for non-jetted NLS1s, the large-scale environment density is low, with most sources living in isolated spiral galaxies\citep{Jarvela17}. On the other hand, while a certain fraction of jetted NLS1s are indeed living in interacting systems \citep{Jarvela18, Berton19a}, just like for high-$z$ sources, this does not seem to be a rule for these objects \citep{Olguiniglesias20, Varglund22, Varglund23}. These results indicate that relativistic jets, or at least jets forming in high-Eddington AGN, do not necessarily form only in the presence of mergers, but also via different triggering mechanisms. Finally, it is worth noting that the discovery of blazars at high-$z$ opened the same question about the parent population that was discussed for jetted NLS1s in Sect. 5.2. The number of jetted AGN in the early Universe, indeed, should be much higher than what we currently observe \citep{Banados25}. Sensitivity limits, especially for unbeamed and lower-power jets, can further increase the number of missing parent sources.

\section{Conclusions}

In this review, we described the physical and observational properties of NLS1s at different frequencies. Our current understanding is that they are low-mass and high-Eddington AGN, possibly representing an early evolutionary stage in active galaxies' life cycle. Therefore, they are excellent laboratories to study in details the physics of AGN in extreme accretion conditions (e.g., central engine properties, outflows, jet launching mechanisms, AGN feedback). However, despite four decades of studies, a lot of properties of NLS1s are still very poorly understood, if not entirely unknown. 

This review effort was done in the light of the discovery of several sources at high-$z$ with the \textit{JWST}, which could be formally classified as NLS1s via their restframe optical spectra. In particular, we examined differences and similarities between NLS1s and low- and high-luminosity high-$z$ AGN in a broader sense. While NLS1s and low-luminosity high-$z$ AGN do share some properties, including their relatively narrow broad-line widths, low black hole mass, high accretion rates (the latter two being model-dependent inferences), and potentially X-ray properties, key differences remain. High-$z$ AGN, and in particular LRDs, appear to be different. The host galaxy-to-SMBH mass ratio is reversed compared to NLS1s, implying different growth histories. Additionally, extreme X-ray weakness in LRDs suggests either an intrinsic suppression of X-ray emission due to accretion physics or observational biases that make these objects difficult to detect. Despite these differences, accretion physics remain the same over the cosmic time, and studying high/super-Eddington accretion in the local Universe can help us shed light on how those same processes were taking place in the early Universe. Future \textit{JWST}, ALMA, and X-ray observations, along with theoretical simulations, will be crucial in further elucidating these connections and refining our understanding of early SMBH growth.

\begin{acknowledgements}
\footnotesize \textit{Acknowledgements.} This review is dedicated to the memory of Alberto Orsan, who participated in the observations of the spectrum of NGC 4051 shown in Fig.~\ref{fig:opt_spectrum}. MB is also grateful to Vincenzo Abate, Andrea Cunial, Soccorsa Selvaggio, and Giacomo Terreran, who all contributed to the observations. We thank S. Panda for providing the datapoints used for the general quasar population in Fig.~\ref{fig:ev1}.
C.M. acknowledges support from Fondecyt Iniciacion grant 11240336 and the ANID BASAL project FB210003. A.T. acknowledges financial support from the Bando Ricerca Fondamentale INAF 2022 Large Grant “Toward an holistic view of the Titans: multi-band observations of $z>6$ QSOs powered by greedy supermassive black holes"
\end{acknowledgements}

\bibliographystyle{aa}
\bibliography{biblio}

\end{document}